\documentclass[aps,prl,showpacs,floatfix,tightenlines,twocolumn,amsmath,amssymb,nobalancelastpage,reprint,superscriptaddress]{revtex4-1}
\usepackage{graphicx}
\usepackage{color}
\usepackage{ulem}
\usepackage{bm}
\usepackage{hyperref}
\usepackage{setspace}
\usepackage{amsmath}
\usepackage{chemformula}
\usepackage{siunitx}
\usepackage{multirow}
\usepackage{booktabs}
\usepackage{dcolumn}
\usepackage{array}

\usepackage{tabularx}

\begin{document}

\title{Local symmetry breaking and orbital glass behaviour in CoFe\textsubscript{2}O\textsubscript{4}}

\author{Soumya Shephalika Behera}
\affiliation{UGC-DAE Consortium for Scientific Research, Indore 452001, India}

\author{Isha}
\affiliation{UGC-DAE Consortium for Scientific Research, Indore 452001, India}

\author{Arvind Kumar Yogi}
\affiliation{UGC-DAE Consortium for Scientific Research, Indore 452001, India}

\author{V R Rao Medicherla}
\email{venkatamedicherla@soa.ac.in}
\affiliation{Department of physics, ITER, Siksha 'O' Anusandhan Deemed to be University, Bhubaneswar 751030, India} 

\author{Parasmani Rajput}
\affiliation{Beamline Development and Application Section, Bhabha Atomic Research Centre ,Trombay, Mumbai 400085, India  } 

\author{Archana Sagdeo}
\affiliation{Accelerator Physics and Synchrotrons Utilization Division, Raja Ramanna Centre for Advanced Technology, Indore 452017, India}
\affiliation{Homi Bhabha National Institute, Training School Complex, Anushakti Nagar, Mumbai 400094, India}

\author{Jaspreet Singh}
\affiliation{Accelerator Physics and Synchrotrons Utilization Division, Raja Ramanna Centre for Advanced Technology, Indore 452017, India}

\author{Vasant Sathe}
\affiliation{UGC-DAE Consortium for Scientific Research, Indore 452001, India}

\author{R J Choudhary}
\email{ram@csr.res.in}
\affiliation{UGC-DAE Consortium for Scientific Research, Indore 452001, India}

\date{\today}

\begin{abstract}

The structural distortions, orbital correlations, and electronic states in cobalt ferrite (CoFe$_2$O$_4$) were investigated using complementary characterisation techniques, including synchrotron X-ray diffraction (SR-XRD), hard X-ray photoelectron spectroscopy (HAXPES), X-ray absorption spectroscopy (XANES), extended X-ray absorption fine structure (EXAFS), and Raman spectroscopy. SR-XRD confirms phase purity and reveals a temperature-dependent superlattice reflection between 200~K and 100~K, consistent with the emergence of short-range orbital ordering driven by cooperative Jahn-Teller distortion (JTD). The disappearance of this feature below 100~K signals orbital freezing and the onset of a glass-like orbital state. HAXPES measurements show multiplet splitting and charge-transfer satellite features in the Co and Fe 2$p$ core levels, indicating mixed valence states and strong electron correlations. XANES analysis reveals hybridized $p$--$d$ states and local coordination distortions. Temperature-dependent EXAFS measurements indicate increasing local disorder---particularly in Fe--O and Fe--Fe octahedral bonds---as evidenced by enhanced Debye--Waller factors. These distortions, attributed to cation redistribution and oxygen vacancies, are static and asymmetric, primarily affecting the octahedral sublattice. Notably, signatures of cooperative Jahn-Teller distortions emerge in the intermediate temperature range ($T_\mathrm{o} \approx 200$--$100$~K) and disappear upon further cooling. Raman spectroscopy further supports these findings, revealing phonon anomalies and enhanced spin-phonon coupling in the same temperature range. Magnetic measurements indicate spin reorientation and exchange interaction anomalies that align with the orbital behaviour. ~Together, these results hint at a frustrated orbital state in CoFe$_2$O$_4$—possibly involving cooperative Jahn-Teller distortions, disrupted long-range coherence, and orbital glass behaviour—offering new insights into the coupling of orbital, spin, and lattice degrees of freedom in spinel systems.

\begin{description}
\item[Keywords]
EXAFS, Debye waller factor, shake-up satellite, magnetoelastic coupling

\end{description}
\end{abstract}

\maketitle


\section{Introduction}  

Transition metal oxides with partially filled $d$-orbitals exhibit a rich  {of electronic, magnetic, and structural  phenomena due to the strong interplay among charge, spin, orbital, and lattice degrees of freedom \cite{tokura2000orbital,li2014direct}. Among these, the orbital degree of freedom-arising from the spatial orientation of degenerate d-orbitals-plays a pivotal role in determining the ground state properties of many correlated systems \cite{lee2012two,li2014direct}.

Orbital ordering, typically stabilized by cooperative JTD, lifts the degeneracy of the e$_g$ orbitals, leading to symmetry-lowering structural transitions and the emergence of long-range ordered phases. This mechanism is well documented in systems such as LaMnO$_3$ \cite{ahmed2009volume,trokiner2013melting,chatterji2003volume}, KCuF$_3$\cite{lee2012two}, and spinel ferrites containing Mn$^{3+}$ and Cu$^{2+}$ ions, where JT distortions are intimately linked to magnetic anisotropy and structural instabilities.
 
\begin{figure*} [t]
\includegraphics[width= 0.9\textwidth]{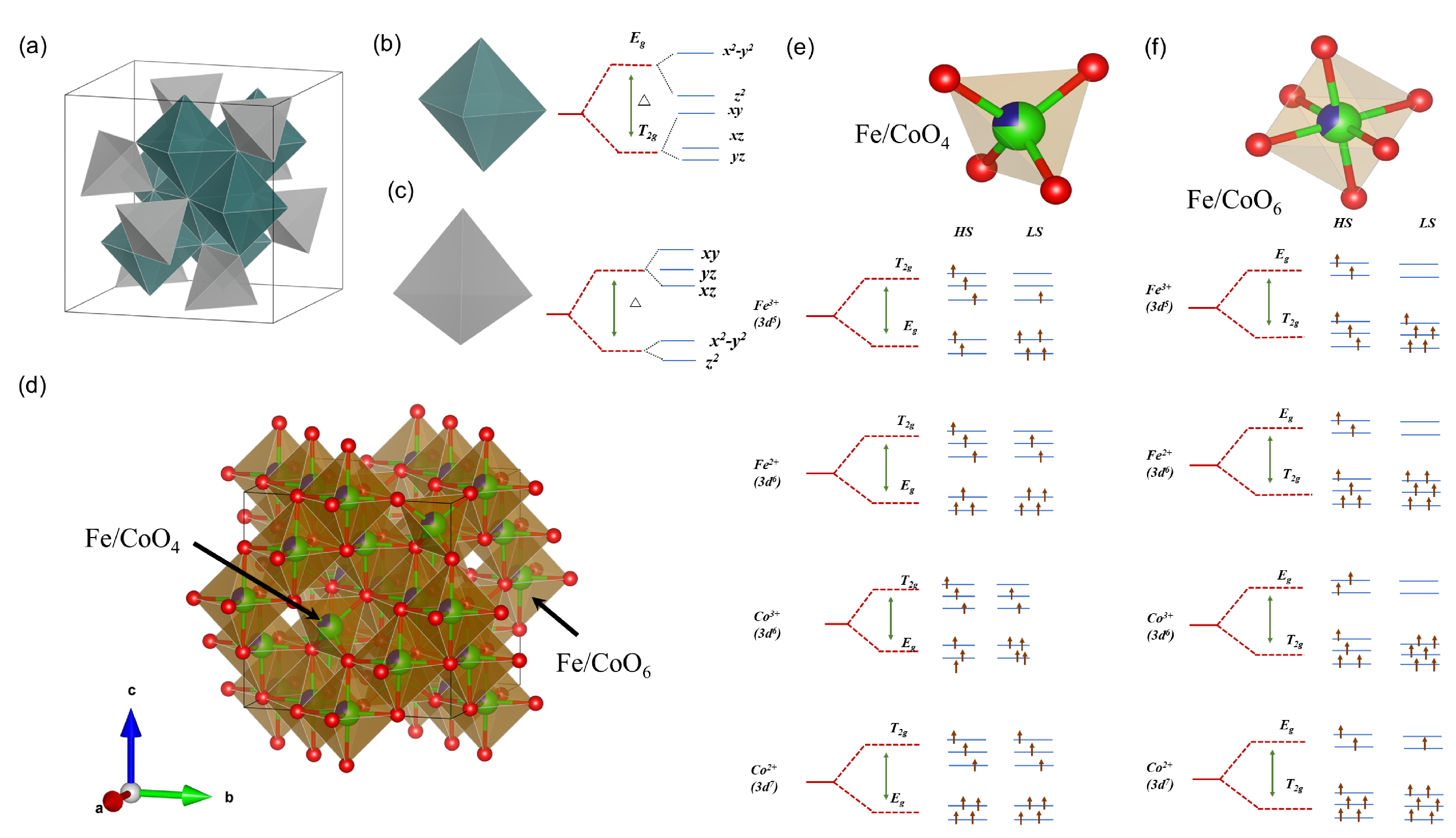}
\caption{\label{fig:wide} \texttt{} (a) Schematic of the spinel lattice showing two distinct crystallographic sites: tetrahedral and octahedral. (b, c) Schematic diagrams of crystal electric field (CEF) splitting and orbital configurations in tetrahedral and octahedral environments. (d) Crystal structure of CoFe$_2$O$_4$ projected onto the $bc$-plane. (e) CEF splitting and orbital degeneracy in the distorted CoO$_4$/FeO$_4$ tetrahedral units and (f) in the CoO$_6$/FeO$_6$ octahedral environment, both influenced by the Jahn–Teller (J–T) effect.}
\end{figure*}

However, in systems with inherent site disorder, geometric frustration, or competing interactions, the development of long-range orbital ordering can be hindered. Instead, such materials may enter a frozen, disordered state known as the orbital glass \cite{tokura2000orbital}. In this state, while local orbital distortions persist due to strong JT coupling, their spatial correlations remain short-ranged and frustrated, preventing global symmetry breaking. The orbital glass state is thus characterized by static but spatially random orbital orientations \cite{tokura2000orbital}, reflecting a delicate balance between ordering tendencies and frustration effects. Understanding the emergence and stabilization of such glassy orbital states is essential for elucidating exotic ground states in correlated oxides.

AB$_2$O$_4$ is a spinel type structure and the magnetic and electrical properties of spinels depend on cations distribution in the tetrahedral (A) and octahedral (B) sublattices of a cubic structure. In particular CoFe$_2$O$_4$ ferrite is a very important spinel type structure with ferrimagnetic behavior (high Curie temperature $T_{c}$ = 793 K), strong magnetocrystalline anisotropy along with a moderate magnetization ($M_s$), and high coercivity ($H_{c}$)~\cite{tokura2000orbital,Yang2008}. Thus, CoFe$_2$O$_4$, a classic inverse spinel ferrite, serves as a compelling system for investigating the interplay between lattice distortions, orbital degrees of freedom, and magnetic behavior. In its inverse spinel structure, Co$^{2+}$ ions predominantly occupy octahedral sites, while Fe$^{3+}$ ions are distributed across both tetrahedral and octahedral positions. The partial cation inversion and competition between crystal field splitting, Hund’s exchange interactions give rise to complex magnetic anisotropy and orbital-lattice coupling. 

The electronic configurations of transition-metal cations in octahedral coordination play a crucial role in driving local structural distortions via the Jahn-Teller (JT) effect. In CoFe$_2$O$_4$, Co$^{2+}$ (3\textit{d$^7$}) and Fe$^{3+}$ (3\textit{d$^5$}) ions occupy octahedral sites in a partially inverse spinel structure. {High-spin Co$^{2+}$ adopts a $t_{2g}^5e_g^2$ configuration and is a JT-active ion, as the system can lower its energy through local lattice distortions that will lift the $t_{2g}$ orbital degeneracy. While Fe$^{3+}$ in the high-spin 3\textit{d$^5$} configuration is typically considered JT-inactive. A schematic of the spinel lattice with tetrahedral and octahedral sites, their crystal field splitting and orbital degeneracy, along with the crystal structure of CoFe$_2$O$_4$ projected onto the \textit{bc}-plane, is shown in Fig.~1(a--e).

The coexistence of JT-active (Co$^{2+}$, Fe$^{2+}$) and non-JT-active (Fe$^{3+}$) ions, coupled with cationic inversion and mixed valence states, leads to a spatially inhomogeneous distortion field. However, the origin of such local structural distortion is still not explained which is crucial to understand it's electronic properties.These distortions remain local and dynamically fluctuating, preventing the development of long-range orbital order. Instead, short-range orbital correlations emerge in the intermediate temperature range\cite{koborinai2016orbital}, which ultimately freeze into a disordered, non-periodic orbital configuration at low temperatures---indicative of an orbital glass state. Such microscopic details for interesting bulk Cobalt-ferrite CoFe$_2$O$_4$ is crucial and we believe that effect of Co and Fe local environment plays an important role in the physical properties which we address by using detailed local probe techniques. For instance EXAFS  render serval important information on interplay between spin state, orbital degeneracy, and local symmetry-breaking arising due to the cooperative JT distortions as we qualitatively explain in Fig. 1. These crucial information about the local bond-distances and the effective coordination numbers of neighboring atoms can help to understand the physical properties of CoFe$_2$O$_4$ in great detail. And thus we design our present study by focusing on local probes techniques such as X-ray absorption spectroscopy (XANES) and extended X-ray absorption fine structure (EXAFS) to capture the local symmetry-breaking which significantly influence the Co and Fe spin states in CoFe\textsubscript{2}O\textsubscript{4}.

In this study, we investigate the structural and electronic properties of CoFe$_2$O$_4$ (CFO), with a particular focus on the role of crystal distortions in shaping its magnetic and electronic behavior. X-ray absorption spectroscopy (XAFS) is employed to probe the local electronic environments of Co and Fe ions, while extended X-ray absorption fine structure (EXAFS) analysis provides insights into their local coordination and atomic structure. The EXAFS data, analyzed using the Debye–Waller factor within the correlated Einstein model, reveal temperature-dependent lattice disorder and the presence of Jahn–Teller distortions (JTD). Hard X-ray photoelectron spectroscopy (HAXPES) measurements at room temperature further confirm the chemical states and mixed valence character of the cations, as well as the presence of charge transfer satellite features, reflecting crystal field effects and electronic correlations.

Synchrotron X-ray diffraction (SR-XRD) is used to examine the long-range crystallographic structure. While the global cubic symmetry remains intact, a distinct superlattice reflection appears between 200~K and 100~K. Complementary Raman spectroscopy supports these findings, revealing a pronounced softening of the E$_g$ phonon mode---consistent with enhanced orbital–lattice coupling and the onset of short-range orbital correlations. Magnetic measurements additionally indicate spin ordering in the same temperature regime, suggesting the emergence of exchange interactions coupled to the orbital \cite{tokura2000orbital}.

\section{Experiments}

Polycrystalline cobalt ferrite CoFe$_2$O$_4$ was synthesized by a conventional solid-state reaction route. The initial ingredients, Fe$_2$O$_3$ and CoO, were ground in stoichiometric amounts to create a homogeneous mixture. The mixture was then subjected to a 12-hour calcination process at \SI{750}{\degreeCelsius} and subsequently calcined at \SI{900}{\degreeCelsius} for 12 hours to synthesize the CoFe$_2$O$_4$ (CFO) sample.\\
 
X-ray Diffraction (XRD), X-ray Absorption Near Edge Structure (XANES), and Extedned X-ray Absorption Fine Structure (EXAFS) as well as Hard X-ray Photoelectron Spectroscopy (HAXPS) measurements were carried out on CoFe$_2$O$_4$ using BL-12, and BL-09 and BL-14 beamlines respectively, at Indus-2 Synchrotron source at Raja Ramanna Center for Advanced Technology (RRCAT), Indore, India. \\

XRD measurements on CoFe$_2$O$_4$ powder were conducted in the temperature range of 50–300 K at the X-ray diffraction beamline (BL-12) of Indus-2 synchrotron utilizing an angle-dispersed setup with a MAR-345 Area detector. A uniform amount of finely grinded CoFe$_2$O$_4$ powder was deposited on Kapton tape and mounted on a liquid helium flow-type cryostat, placed on the Image plate sample holder setup. The diffraction geometry, including the sample-to-detector distance ($d = 200$ mm) and the perpendicularity of the detector to the X-ray beam, were calibrated by refining the diffractograms of a reference LaB$_6$ sample, with an estimated wavelength of 0.63658 \AA. The 2D diffractograms were integrated using the FIT2D software to generate standard intensity versus $2\theta$ 1D diffractograms. Full Debye ring integration from the 2D diffractograms ensured a high signal-to-noise ratio, which is particularly effective in revealing parasitic phases and minimizing the effects of privileged orientations, making the data ideal for structure refinement. Crystallographic structural parameters as a function of temperature were obtained by refining the 1D patterns using the full-profile Rietveld method in FULLPROF software.\\

The core levels of Co and Fe in CFO have been investigated using Hard X-ray Photo-Electron Spectroscopy (HAXPES) employing a photon energy of 4000 eV at beamline BL14 of the Indus-2 synchrotron. The vacuum in the analysis chamber was maintained in the \(10^{-10}\) Torr range. The total resolution in the measurement was set to 0.8 eV to obtain a good signal-to-noise ratio. The surface cleaning procedures were purposefully avoided to prevent any surface damage that may induce extra features in photoemission.The binding energy of the photoelectrons was calibrated using 1s core level binding energy of contaminant carbon at 284.6 eV. This procedure for binding energy reference automatically compensates for any shift in binding energy due to the static charging of the sample during photoemission. During the measurement, no peak shift was observed between different scans, suggesting the absence of dynamic charging of the sample.\\

X-ray absorption spectroscopic data were recorded at the Fe and Co K-edges in transmission mode for all samples using the BL-09 beamline (4–25 keV) at the Indus-2 synchrotron source. The polycrystalline powder samples were mixed thoroughly with boron nitride (BN) using a mortar and pestle in proportions, ensuring that the absorption edge jump ($\Delta\mu$) ranged from 0.7 to 1, as calculated using the XAFSMASS code \cite{klementiev2016xafsmass}. X-ray absorption fine-structure (XAFS) spectroscopy experiments were carried out in TEY mode using hard X-ray synchrotron radiation using Athena and Artemis software and  X-ray absorption near-edge structure (XANES) and extended X-ray absorption fine-structure (EXAFS) spectral measurements in the energy range of 5 to 25 keV using BL-09 beamline of Indus-2 synchrotron. \\

Magnetization measurements were carried out with the help of the MPMS 7 T SQUID-VSM instrument (Quantum Design Inc., USA). Raman measurements were carried out using a Horiba JY HR-800 spectrometer equipped with excitation sources of 473 nm and 633 nm. The system employs an 1800 grooves/mm grating for spectral dispersion and a charge-coupled device detector for signal acquisition. Additionally, an external setup was integrated to enable low-temperature measurements using liquid nitrogen up to 80K.\\

\begin{figure*} [t]
\includegraphics[width= 1\textwidth]{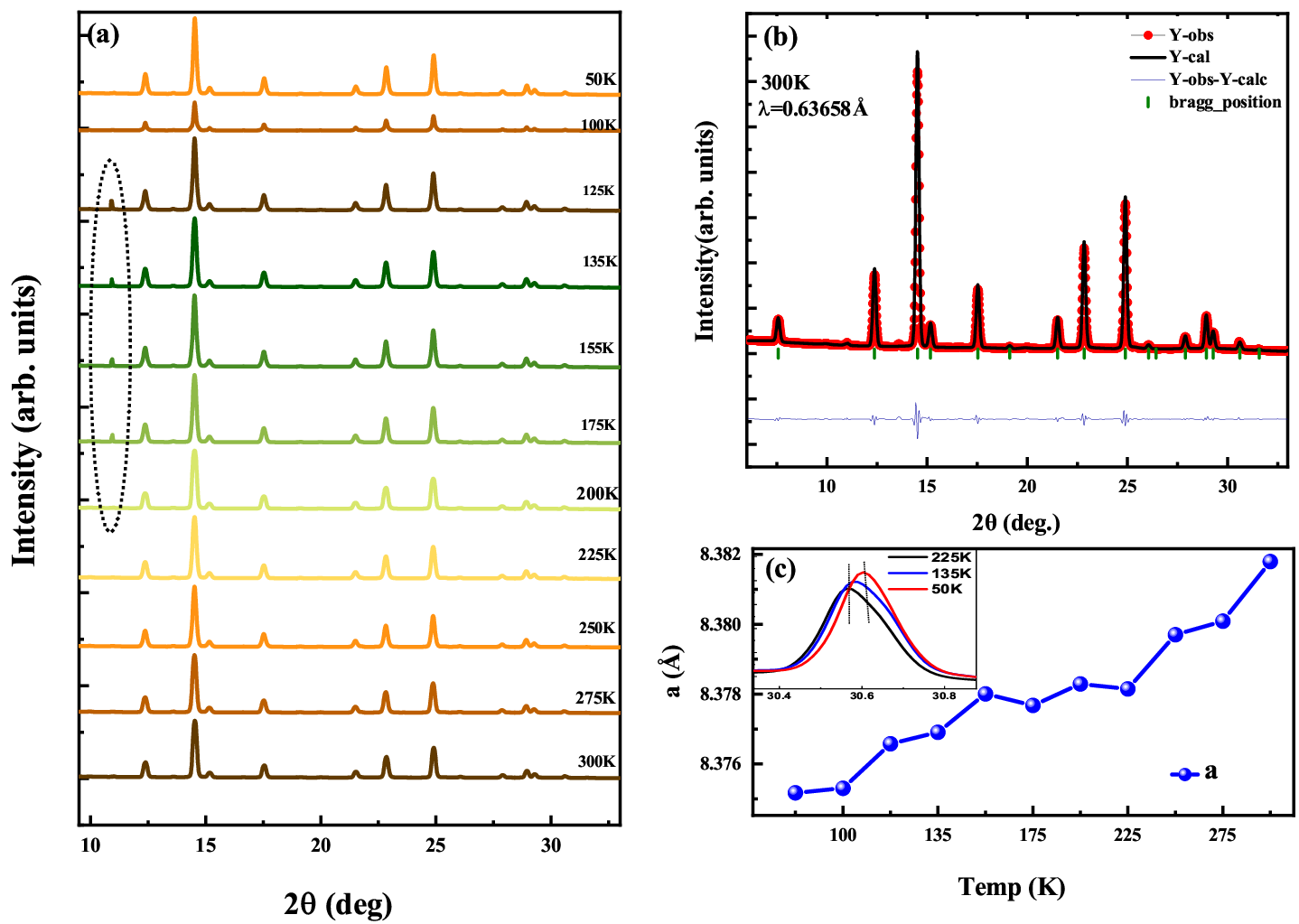}
\caption{\label{fig:wide} \texttt{} (a) XRD patterns of CoFe\textsubscript{2}O\textsubscript{4} measured in the temperature range of 50–300 K; (b) Rietveld refinement results of the XRD pattern at 300 K; (c) Variation of lattice parameter as a function of temperature with the inset showing the temperature-dependent shift in peak position for selected temperatures (50K, 135K, 225K).}
\end{figure*}

\begin{figure} [h]
\includegraphics[width= 0.5\textwidth]{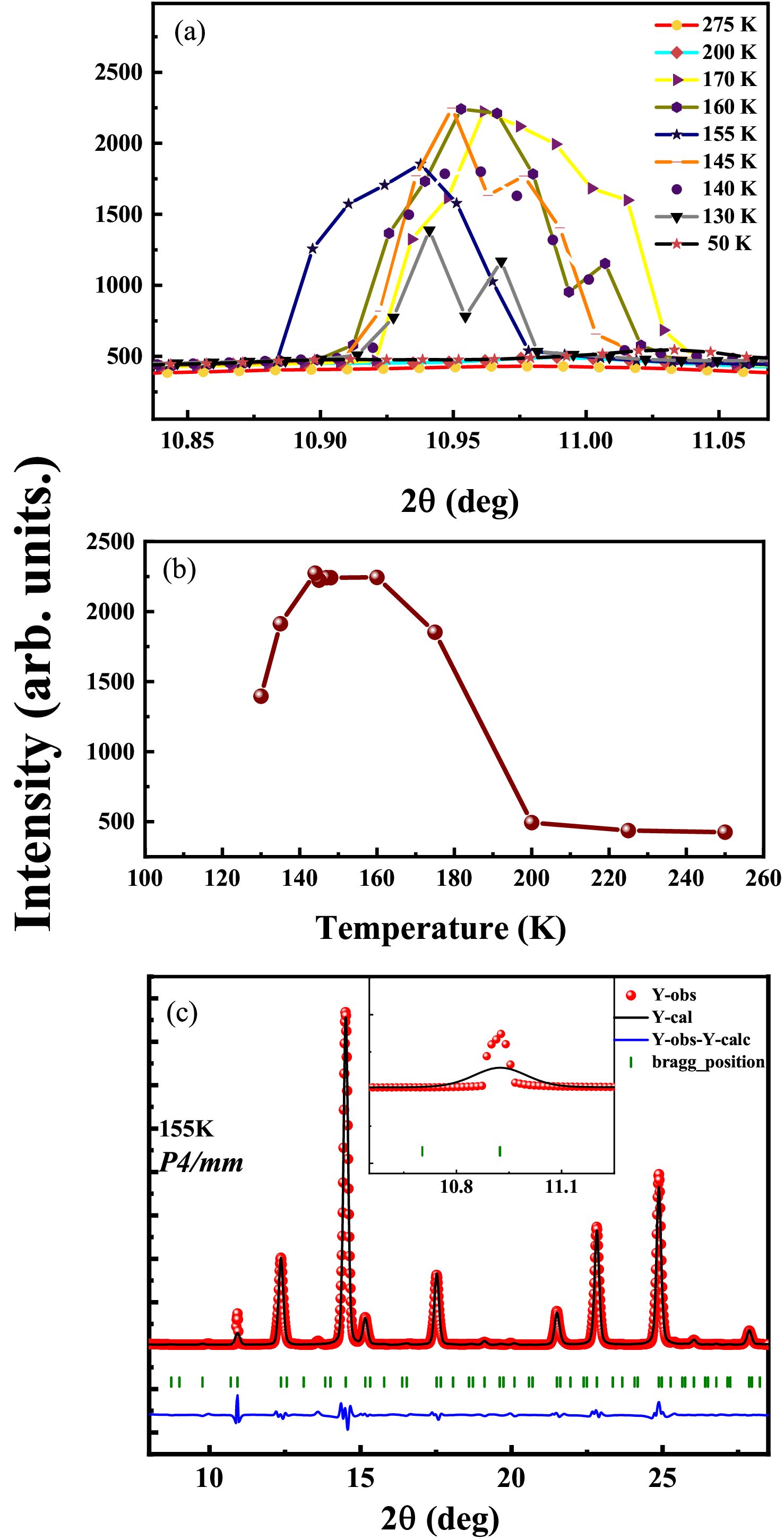}
\caption{\label{fig:wide} \texttt{} (a) Temperature evolution of the superlattice peak at \(2\theta \approx 10.95^\circ\). (b) Temperature dependence of its intensity. (c) Le Bail fit at 155\,K using \(P4/mm\) symmetry; inset shows the superlattice peak and Bragg position.}
\end{figure}

\begin{figure} [h]
\includegraphics[width= 0.5\textwidth]{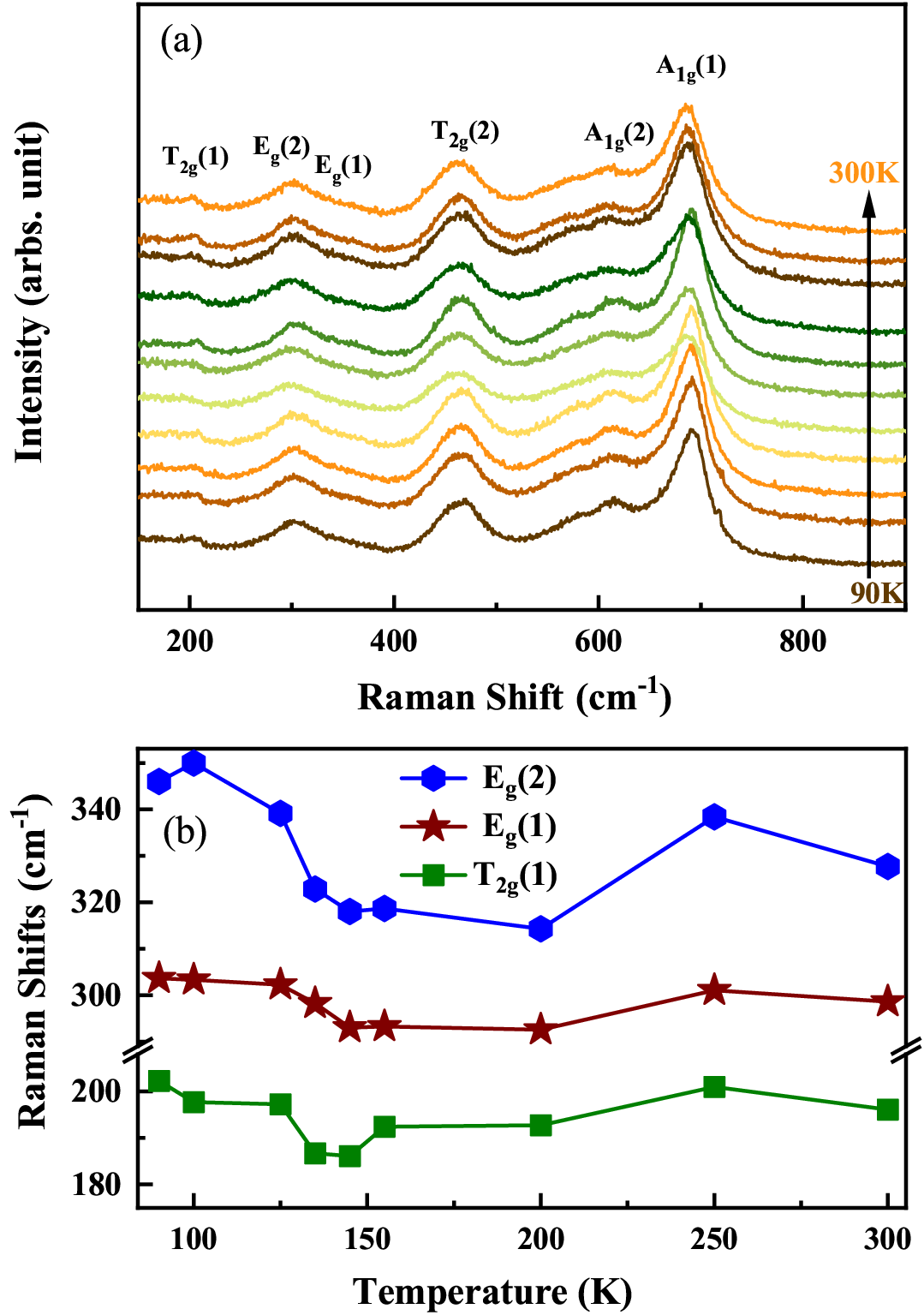}
\caption{\label{fig:wide} \texttt{}(a) Raman spectra recorded over the temperature range of 90K to 300K, showing intensity variations with Raman shift. (b) Temperature dependence of Raman shifts for \(T_{2g}(1)\), \(E_g(1)\), and \(E_g(2)\) modes.}
\end{figure}

\begin{figure} [h]
\includegraphics[width= 0.5\textwidth]{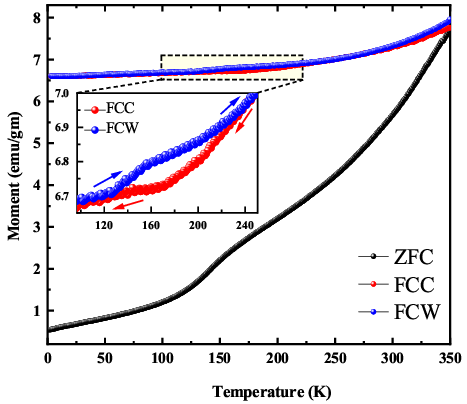}
\caption{\label{fig:wide} \texttt{} Magnetization data of CoFe\textsubscript{2}O\textsubscript{4}, showing the temperature dependence of magnetization measured under zero-field-cooled (ZFC), field-cooled warming (FCW), and field-cooled cooling (FCC) conditions. The main plot highlights the magnetic behavior and transitions across the temperature range, while the inset illustrates the bifurcation between the FCC and FCW curves.
}
\end{figure}

\section{Results and Discussion}

\subsection{Temperature-dependent XRD}

To evaluate possibility of any structural symmetry-breaking transition in CoFe$_2$O$_4$, we performed Temperature-dependent synchrotron X-ray diffraction (SR-XRD) measurements in the 300--50~K range. The detailed structural analysis by SR-XRD shows drastic change between 200~K and 100~K as shown in Fig. 2 (a). Temperature-dependent SR-XRD structural data analysis and the bond valence sum (BVS) calculation are performed by FullProf software~\cite{rodriguez1993recent}. The Rietveld refinement of one SR-XRD pattern at the room temperature (300 K) by the cubic $Fd-3m$ space group is shown in Fig.~ 2 (b). The lattice parameters at different temperatures were obtained by similar Rietveld refinement of the XRD patterns at different temperatures and were plotted in Fig.~ 2(c), and the insert showing the peak shifts progressively to higher angles with decreasing temperature, confirming lattice contraction. 

Moreover, upon cooling, a distinct weak but pronounced peak possibly a super-lattice reflection, as shown in highlighted dashed line in Fig.~3 (a). It appears at 2$\theta$ = 10.95 degree, just below $\sim$200~K, indicating the emergence of super-structure or lattice modulation in CFO. Interestingly the intensity of this peak increases with lowering temperature up to $\sim$100~K, then it starts decreasing as shown in Fig.~ 3(b), possibly due to the development of strong short-range orbital correlations. Such orbital correlations may also break the symmetry and therefore our detailed temperature dependent SR-XRD suggesting a possible structural symmetry-breaking transition does exist in the temperature  200-100 K. It is in line with other measurements, as most of the physical and local probe low-temperature experiments on this compound show a clear change in this temperature interval only. Further, we try to include this super-lattice peak in our SR-XRD refinement as well under various lower symmetries of the inverse spinel CoFe$_2$O$_4$ structure by trial and error method for the automatic indexing of Powder diffraction patterns using DICVOL04 option of Fullprof software~\cite{rodriguez1993recent}. However, from our various attempts we could not successfully index the peak, most probably due to powder XRD avaraging effect. X-ray diffraction of single-crystal thus would become crucial to explain this new super-lattice peak. From our try we at least found that the super-structure weak Bragg peak emerging at ~$2\theta$ = 10.95 degree can be explained by $P4/mm$ symmetry, a simulated pattern with $P4/mm$ from LeBail fit refinement is shown in Fig. 3 (c) and inset shows the super-lattice peak along with Bragg position. However, attempt to fully fit this super-lattice peak (2$\theta$ = 10.95 degree) through either profile (LeBail-fit) or Rietveld refinement was failed [see inset of Fig. 3 (c)] and thus here we explain the emergence of this new super-lattice peak (2$\theta$ = 10.95 degree) by an average structure analysis. This $P4/mm$ symmetry nicely explained rest all of the fundamental Bragg peaks as well [see Fig. 3 (c)], very similar to all the Bragg fundamental reflections as indexed by assuming the cubic space group $Fd-3m$, where all the crystallographic sites were fully occupied. However, with cubic space group $Fd-3m$ we found a drastic change in the super-structure region. The refined lattice parameters for high temperature phase with cubic space group $Fd-3m$ (no super structure) at 300 K are found to be $a = 8.381797 (13)$~\r A. However, at 155 K, without including super-structure, and considering only fundamental reflections, the lattice parameters are found to be $a = 8.3780037 (16)$~\r A. The crystal structure of CoFe$_2$O$_4$ for high temperature phase is shown in Fig. 1 (d). Moreover, an anomaly in the temperature dependent lattice parameter is observed in the temperature range of 100 to 200 K, as shown in Fig.~ 2 (c). Notably, Rietveld refinement confirms that the global symmetry remains cubic throughout, implying that the modulation originates from local JTD of the octahedra, rather than a long-range structural phase transition.

Below 100~K, the superlattice peak abruptly disappears \cite{tokura2000orbital}, reflecting the breakdown of coherent orbital correlations. Instead of stabilizing into a long-range orbitally ordered phase, the system enters a frustrated, disordered regime where dynamic JT distortions freeze into a static configuration. The persistence of cubic symmetry and the disappearance of orbital modulation at low temperatures point toward the formation of an orbital glass state, characterised by randomly frozen orbital orientations and suppressed long-range order. These findings highlight the delicate interplay between orbital-lattice coupling and frustration, which can favour glassy freezing over conventional orbital ordering. To further investigate this phenomenon, we perform temperature-dependent Raman measurements, as presented below.\\

\begin{figure} [h]
\includegraphics[width= 0.5\textwidth]{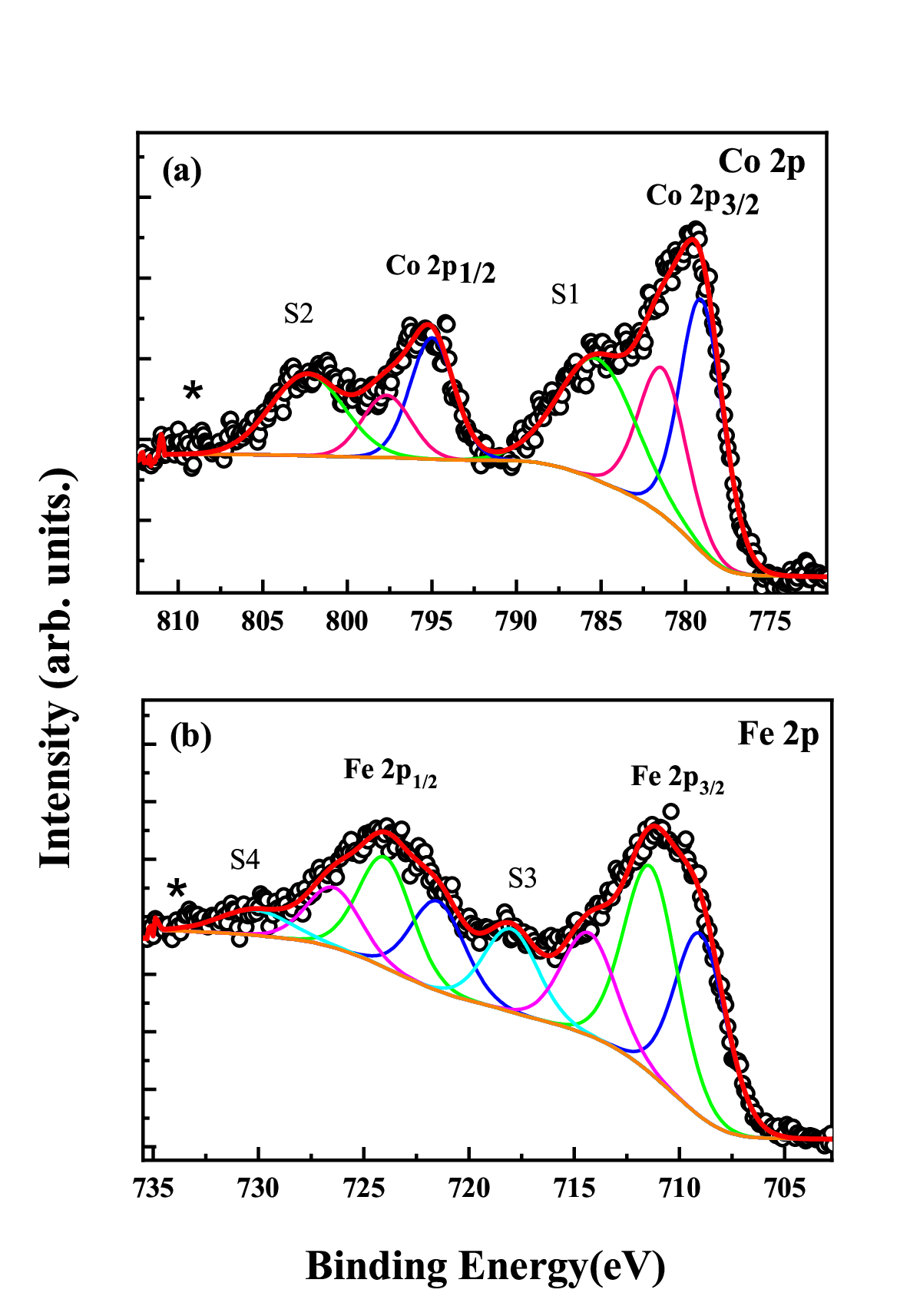}
\caption{\label{fig:wide} \texttt{} The 2p core level of (a) Co and (b) Fe excited by 4 keV synchrotron radiation.
}
\end{figure}

\begin{figure*} [t]
\includegraphics[width= 1\textwidth]{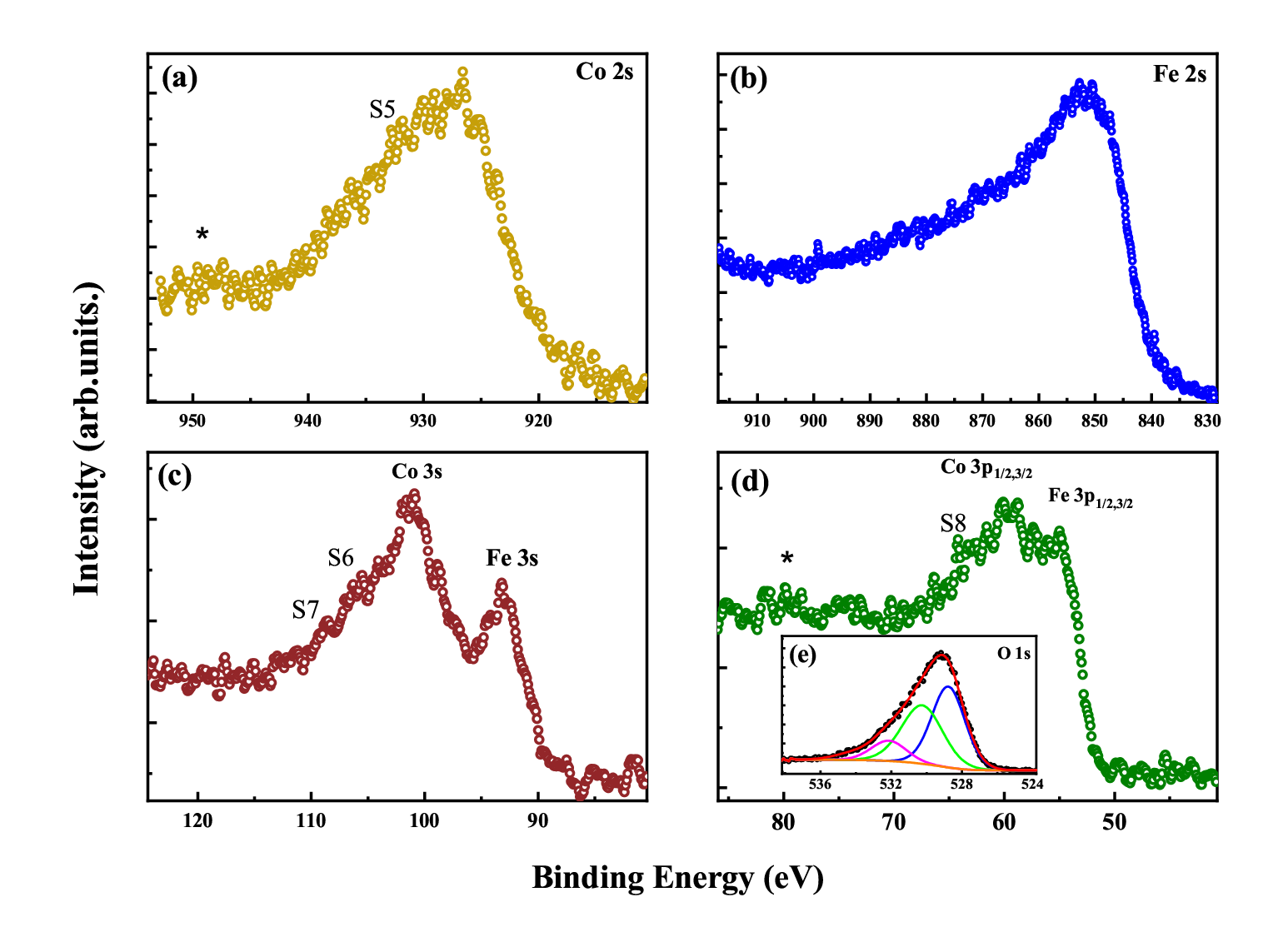}
\caption{\label{fig:wide} \texttt{} Core-level spectra excited by 4 keV synchrotron radiation: (a) Co 2s, (b) Fe 2s, (c) Co and Fe 3s, and (d) Co and Fe 3p. The inset in (d) presents the corresponding (e) O 1s spectrum.
}
\end{figure*}

\begin{figure} [h]
\includegraphics[width= 0.5\textwidth]{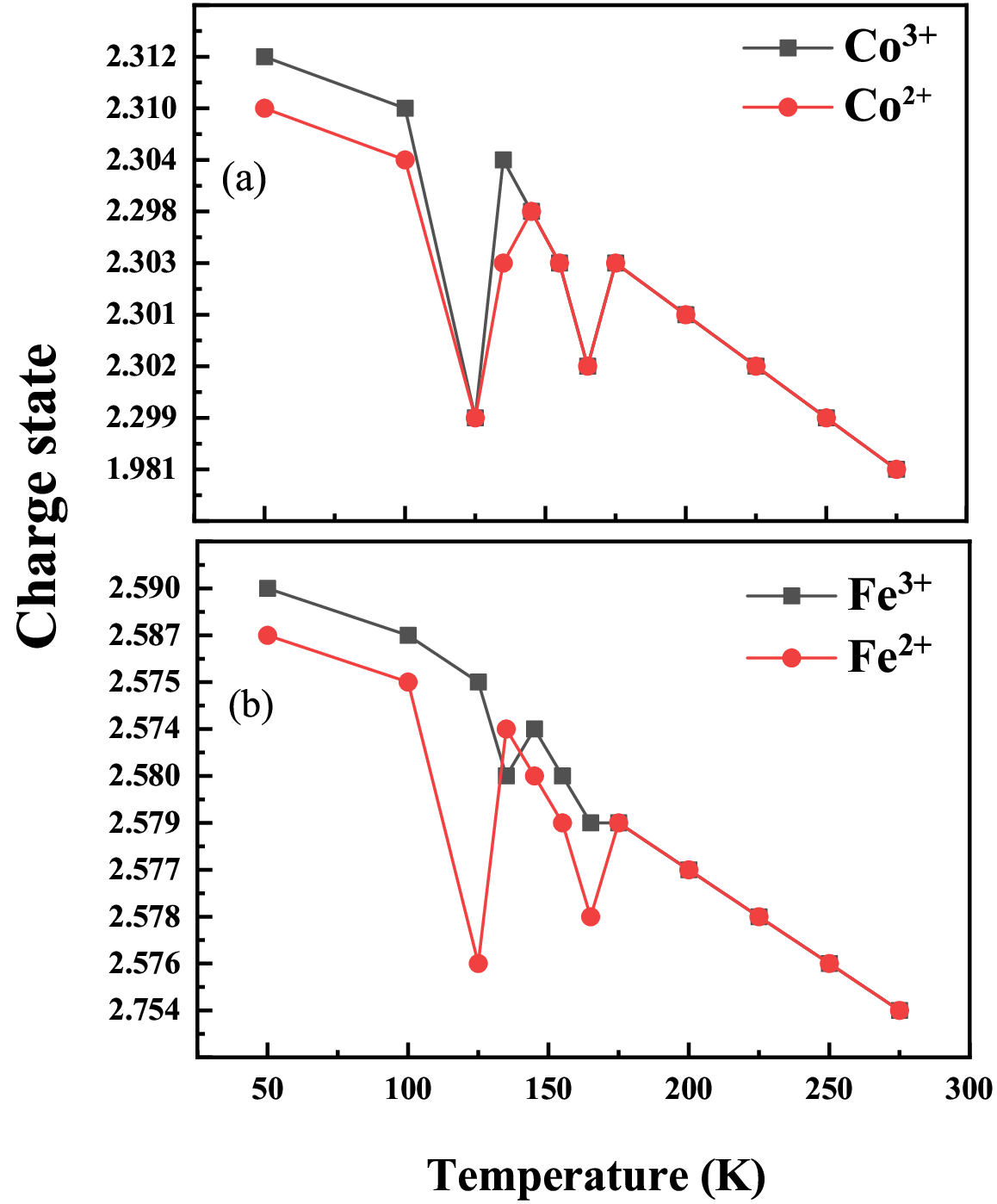}
\caption{\label{fig:wide} \texttt{} Temperature-dependent bond valence sum (BVS) of (a) Co and (b) Fe ions in CoFe$_2$O$_4$.}
\end{figure}

\begin{figure} [t]
\includegraphics[width= 0.5\textwidth]{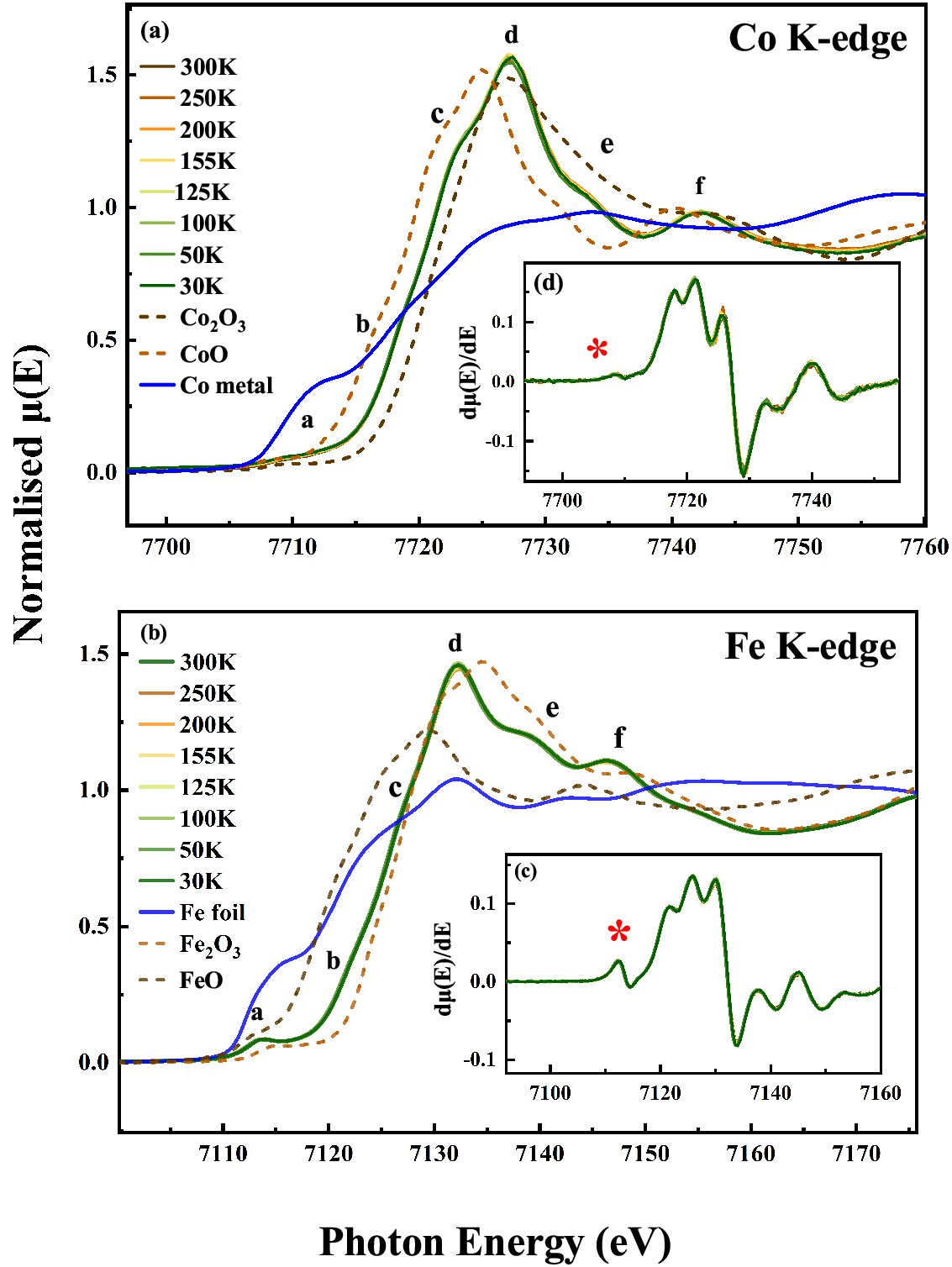}
\caption{\label{fig:wide} \texttt{}(a) Normalized XANES spectra of CoFe\textsubscript{2}O\textsubscript{4}, highlighting the main absorption region of Co K-edge (b) Normalized Fe K-edge XANES spectra. (c) and (d) display the first derivative of the experimental absorption spectra for the Co K-edge and Fe K-edge, respectively.}
\end{figure}

\subsection{Raman Spectroscopy}

Raman spectroscopy is a powerful tool for probing lattice dynamics and local structural changes in systems where spin, orbital, and lattice degrees of freedom are intertwined. In CoFe$_2$O$_4$ (CFO), deviations from the ideal spinel structure often activate additional Raman modes and lead to peak splitting, typically caused by crystal field effects, JTD, and cation disorder. These effects are amplified in inverse spinels due to mixed cation occupancy, making Raman spectroscopy sensitive to subtle structural and electronic changes.

Factor group analysis predicts that the ideal spinel structure with space group \textit{$Fd-3m$} exhibits five first-order Raman-active modes: A$_{1g}$ + E$_g$ + 3T$_{2g}$. In mixed spinels, a significant increase in Raman-active modes is typically observed due to symmetry lowering caused by structural distortions~\cite{almutairi2024unveiling}. The temperature-dependent Raman spectra of CFO reveal distinct phonon shifts which highlight the interplay between lattice dynamics, spin-phonon coupling, and charge redistribution, as shown in Fig.~4. At high temperatures (300--200~K), the T$_{2g}$(1) mode (Fe$^{3+}$--O bending, octahedral) exhibits slight softening from 196.1~cm$^{-1}$ to 192.7~cm$^{-1}$, reflecting thermal expansion. The E$_g$ mode, involving mixed tetrahedral-octahedral vibrations, also softens. The mode at 565.4~cm$^{-1}$ observed at 300~K begins to soften below 200~K, indicating the onset of changes in magnetic arrangement likely associated with spin-phonon coupling \cite{almutairi2023unraveling,xu2007orbital,sun2012spectroscopic}. Additionally, the A$_{1g}$ mode (tetrahedral Fe--O stretching) shifts from 658.3~cm$^{-1}$ to 665.6~cm$^{-1}$, likely due to cation redistribution \cite{yadav2017impact,behera2017influence,chandramohan2011cation}. As the temperature further decreases to the 200--100~K range, phonon hardening becomes evident, particularly in the A$_{1g}$ mode (from 690.4~cm$^{-1}$ to 693.1~cm$^{-1}$) and T$_{2g}$(3) (from 614.2~cm$^{-1}$ to 617.9~cm$^{-1}$), indicating increased lattice stiffness and enhanced magnetoelastic coupling. The A$_{1g}$(2) mode also shows unexpected softening around 145~K.

These phonon anomalies are indicative of evolving orbital-lattice interactions and possible local structural rearrangements. In particular, the pronounced softening of the E$_g$(2) mode and the shift in spin-phonon coupling below 200~K suggest the emergence of short-range orbital correlations and distortions. This behavior is characteristic of JTD, inducing local lattice distortions \cite{tokura2000orbital}. Below 100~K, the hardening and stabilization of the A$_{1g}$ and E$_g$ modes indicate that orbital fluctuations have frozen out \cite{lee2012two}, resulting in a disordered orbital configuration. This is consistent with a frozen orbital glass state, in which local lattice distortions persist without a macroscopic change in symmetry. The presence of additional Raman modes across all temperatures further suggests persistent structural distortion \cite{nandan2019cation,almutairi2024unveiling,almutairi2023unraveling}, and cationic disorder, all of which contribute to the complex vibrational behaviour observed in CFO. The influence of magnetic ordering on Raman mode can be evidenced clearly from the magnetisation data described below.

\subsection{Magnetisation}

The temperature-dependent magnetisation curves of CFO (Fig.~5) reveal a clear thermal hysteresis between field-cooled cooling (FCC) and warming (FCW) in the 200--100~K range, indicating a first-order-like transition and also suggesting magnetic metastability and frustration, primarily arising from the interplay of antiferromagnetic and ferromagnetic exchange interactions between tetrahedral (A-site) and octahedral (B-site) cations via oxygen. It should be noted that in this temperature range, we also observed superlattice reflection in XRD, anomaly in lattice parameter, and Raman behaviour. These occurrences indicate towards coupling between spin and degrees of freedom.In this temperature window, JTD at octahedral sites may partially lift orbital degeneracy, modulating the magnetic exchange interactions \cite{tokura2000orbital}. The resulting spin frustration and orbital-lattice coupling can lead to short-range orbital correlations \cite{tokura2000orbital}. These effects are consistent with the Kugel-Khomskii model\cite{kugel1973crystal,koborinai2016orbital}, where orbital occupancy influences magnetic exchange pathways. Similar phenomena have been reported in manganites and vanadates. Below 100~K, the convergence of FCC and FCW curves and the hardening of Raman modes point to a frozen orbital configuration, suggestive of an lattice coupling, with the observed hysteresis linked to fluctuating orbital states \cite{lee2012two} and spin reorientation.

The interplay between antiferromagnetic and ferromagnetic interactions becomes more pronounced at temperatures in between 200K to 100K. The observed changes suggest a crucial role of A-B superexchange interactions, where Fe$^{3+}$-O-Fe$^{2+}$ and Co$^{2+}$-O-Co$^{3+}$ charge fluctuations influence the overall magnetic structure. This interplay leads to spin frustration effects and local magnetic ordering, which are also reflected in the temperature-dependent Raman shifts and broadening of phonon modes.

\subsection{Hard X-ray photoemission spectroscopy}

To examine the chemical valence states of Co and Fe in CoFe\(_2\)O\(_4\) (CFO), the core-level spectra were analysed using X-ray photoelectron spectroscopy (XPS). Figure- 6 illustrates the Co and Fe 2p core-level spectra of CFO  using 4 keV photon energy.  Both Co 2p and Fe 2p spectra are complex and exhibit multiple features as seen. The spectra were numerically deconvoluted using Gaussian and Lorentzian line shape functions. The background was meticulously fitted using the Shirley method, ensuring precise determination of peak positions and relative intensities for a comprehensive analysis. The Co and Fe \(2p\) core-level spectra, split into \(2p_{1/2}\) and \(2p_{3/2}\) components due to spin-orbit coupling, as indicated in the  Fig.~6. The asymmetry and broadening of the spectral features indicate the presence of multiple valence states for both Co and Fe. Considering this asymmetry and broadening, the Co 2p XPS spectrum has been deconvoluted into distinct peaks centered at 779.0~eV (\(\text{Co}^{2+}\, 2p_{3/2}\)), 781.3~eV (\(\text{Co}^{3+}\, 2p_{3/2}\)), 794.9~eV (\(\text{Co}^{2+}\, 2p_{1/2}\)), and 797.6~eV (\(\text{Co}^{3+}\, 2p_{1/2}\)). Additionally, two satellite features are observed at 785.20~eV and 802.36~eV, respectively \cite{saha2023electronic,kim1975Co0}.The presence of \(\text{Co}^{3+}\) ions in the sample is attributed to cation inversion or Co antisite defects, involving the exchange of Fe and Co ions between the A and B sites. Specifically, we assume that Co ions occupying the A site exist in the \(\text{Co}^{3+}\) oxidation state, while Fe ions migrating from the A site to the B site retain their \(\text{Fe}^{3+}\) state \cite{kobayashi2022intervalence}.

The Fe 2p XPS spectrum, considering its asymmetry and broadening, has been deconvoluted into distinct peaks at 709 eV (Fe$^{2+}$ 2p$_{3/2}$), 711.3 eV (Fe$^{3+}$ 2p$_{3/2}$), 721.4 eV (Fe$^{2+}$ 2p$_{1/2}$), and 723.9 eV (Fe$^{3+}$ 2p$_{1/2}$). Additionally, the satellite features have been fitted with four peaks at 714.3 eV, 718.0 eV, 726.4 eV, and 730.1 eV, respectively\cite{bagus2021combined}.

The 2s and 3s,3p core-level states of Co and Fe are discussed below. The binding energies of Fe's 3s and 3p levels overlap significantly with those of Co, making it somewhat difficult to clearly describe the satellite features of Fe in these levels. For Co 2s, the main peak appears around 927.3 eV, while the satellite feature, denoted as S5, arises at approximately 932.6 eV. The separation between the main peak and the satellite peak in the Co 2s shell is about 5.3–5.6 eV, which is too large to be explained by multiplet splitting \cite{kim1975Co0}. Instead, the satellite peak S5 is attributed to an O 2p~$\rightarrow$~Co 3d shakeup process. Similarly, the satellite structures (peaks S1 and S2) observed in the Co 2p\textsubscript{3/2} and 2p\textsubscript{1/2} regions are also assigned to the same shakeup process, as they exhibit similarities to peak S5 in the Co 2s region in terms of both intensity and location. The unusually broad widths of the main peak and S5 in the Co 2s spectrum are thought to result from multiplet splitting in addition to Coster-Kronig processes.

In the Co 3s photoemission spectrum, the main peak is observed at 101.2 eV, with two satellite peaks, S6 and S7, appearing at 106.3 eV and 112.3 eV, respectively. The complex structure of the Co 3s peak is primarily due to multiplet splitting, which occurs when different electronic states interact and split into multiple peaks. However, the strong satellite features in the Co 2p and 2s regions suggest that peak S6 is caused by both multiplet splitting and a shakeup process, where an electron is ejected and another electron is excited to a higher energy level. The lower binding-energy side of peak S6 is attributed to multiplet splitting, while the higher binding-energy side is due to shakeup. Peak S7 is related to the shakeup process connected to the multiplet splitting in peak S6. To fully understand the Co 3s spectrum, detailed theoretical calculations are needed to explain the interaction of these effects. In the Co 3p region, the satellite structure is less clearly defined compared to the Co 3s region. The main component of S8 is thought to originate from multiplet splitting. Weak satellite structures, about 20 eV below the main peaks in the Co 2s, 2p, 3s, and 3p regions, are denoted with a symbol  (*) and are associated with characteristic energy loss.
The observed multiplet splitting energy in Co is $\Delta E_{3s} \sim 5.1 $\, \text{eV}.

In the Fe 2s core-level spectrum, the main peak is observed around 851.7 eV \cite{kuvcas2019evolution}. The region around this peak is broad, but no additional satellites is clearly visible because of low signal-to-noise ratio \cite{kim1974charge}. The broadening of the main peak may be due to effects such as multiplet splitting or broadening from other electronic interactions, but without clear satellite peaks, the interpretation is less complex compared to the other core levels.In the Fe 3p core-level spectrum, the main peak is observed around 55.9 eV\cite{fujimori1986photoemission}. However, at higher binding energies, the Fe 3p spectrum merges with the Co 3p spectrum, making it difficult to analyze or identify any distinct satellite features for Fe 3p. Similarly, in the Fe 3s spectrum, the main peak appears at approximately 93.5 eV, but its higher binding energy side merges with the Co 3s peak. Due to these overlaps and the merging of spectral features, it is not possible to clearly define or explain the satellite structures in the Fe core-level spectra.Deconvolution of the O\,1\textit{s} XPS spectra in insert Figure~7(d) reveals three main peaks at binding energies of 529.8, 531.5, and 533.3~eV, corresponding to lattice oxygen (O$_\mathrm{L}$), oxygen vacancies or defects (O$_\mathrm{V}$), and chemisorbed oxygen species (O$_\mathrm{C}$), respectively\cite{majumder2019elucidating}

The chemical valency analysis of the CFO samples, based on the integrated intensities of the fitted XPS peaks, reveals a mixed-valence state for both Co and Fe ions. The results indicate that cobalt is present as approximately 65\% Co$^{2+}$ and 35\% Co$^{3+}$, while iron shows a distribution of about 44.7\% Fe$^{2+}$ and 55.3\% Fe$^{3+}$.

\begin{figure} [h]
\includegraphics[width= 0.5\textwidth]{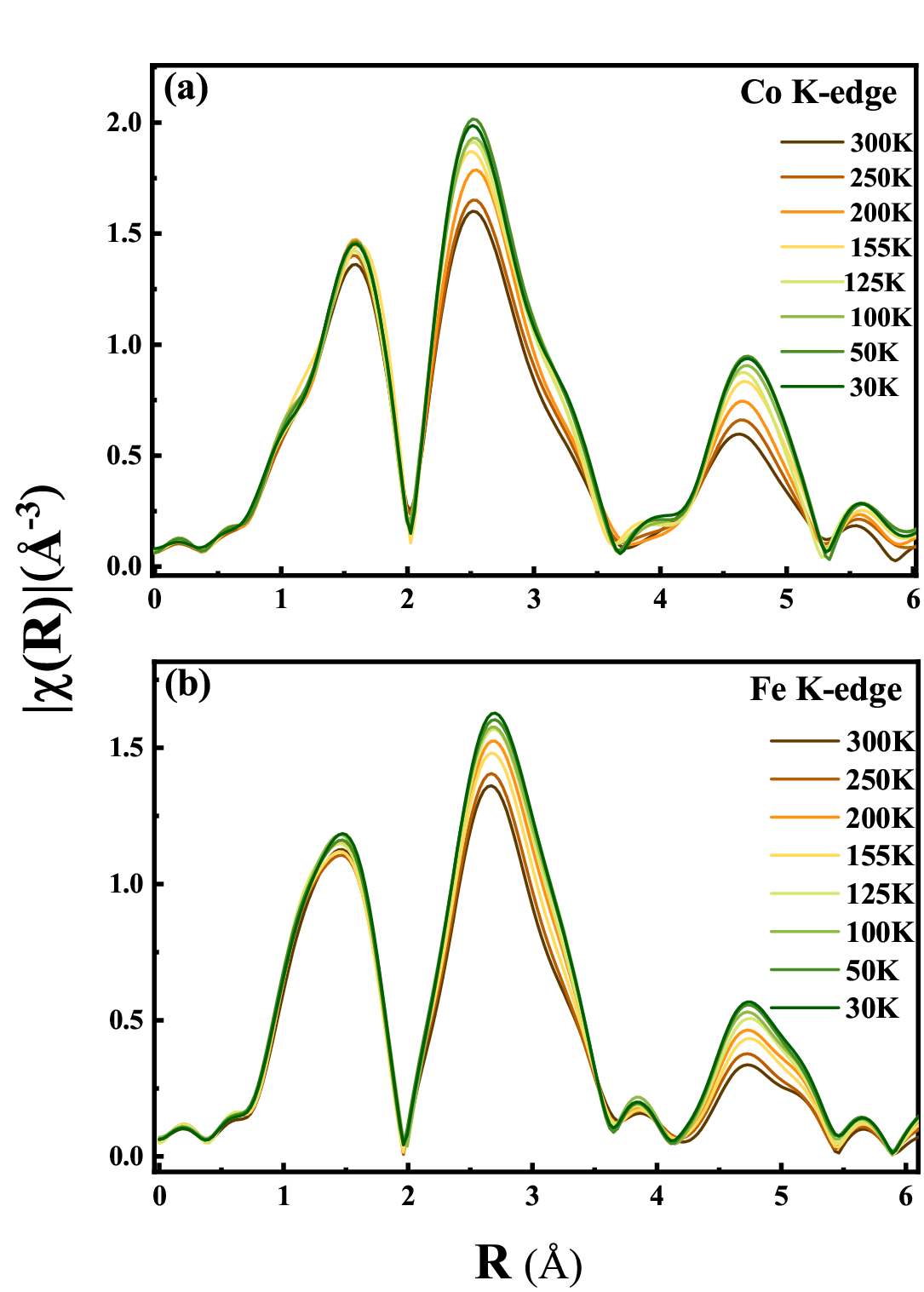}
\caption{\label{fig:wide} \texttt{}}The temperature-dependent EXAFS data: (a) Co K-edge, (b) Fe K-edge.
\end{figure}

\begin{figure} [h]
\includegraphics[width= 0.5\textwidth]{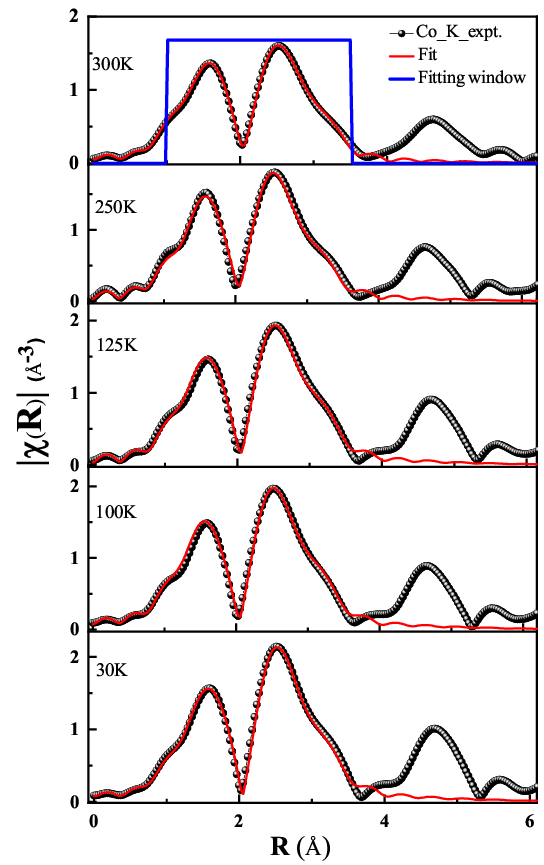}
\caption{\label{fig:wide} \texttt{}Fitting of the $|\chi(R)|$ spectra at the Co K-edge in R-space. The modulus of the Fourier transform ($|FT|$) of $k^3$-weighted Fe EXAFS data is shown along with the best-fit curves. The data is displayed for a range of temperatures.}
\end{figure}

\begin{figure} [h]
\includegraphics[width= 0.5\textwidth]{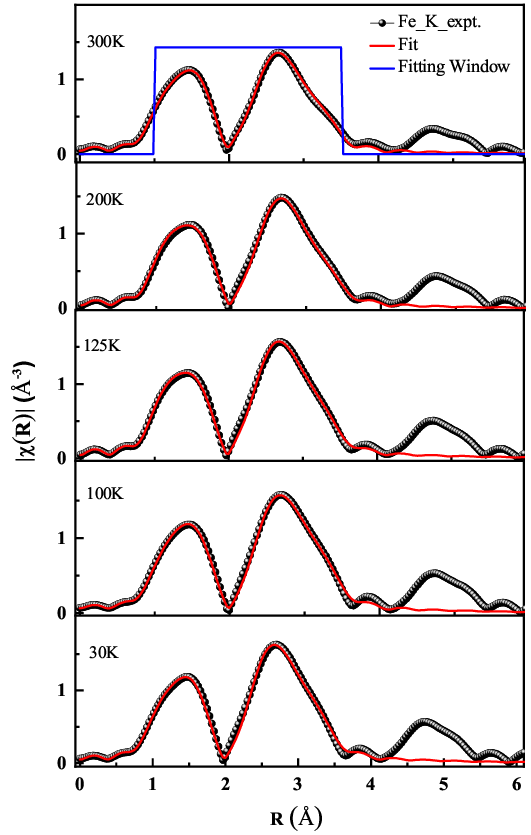}
\caption{\label{fig:wide} \texttt{}Fitting of the $|\chi(R)|$ spectra at the Fe K-edge in R-space. The modulus of the Fourier transform ($|FT|$) of $k^3$-weighted Fe EXAFS data is shown along with the best-fit curves. The data is displayed for a range of temperatures.}
\end{figure}

\begin{figure} [h]
\includegraphics[width= 0.5\textwidth]{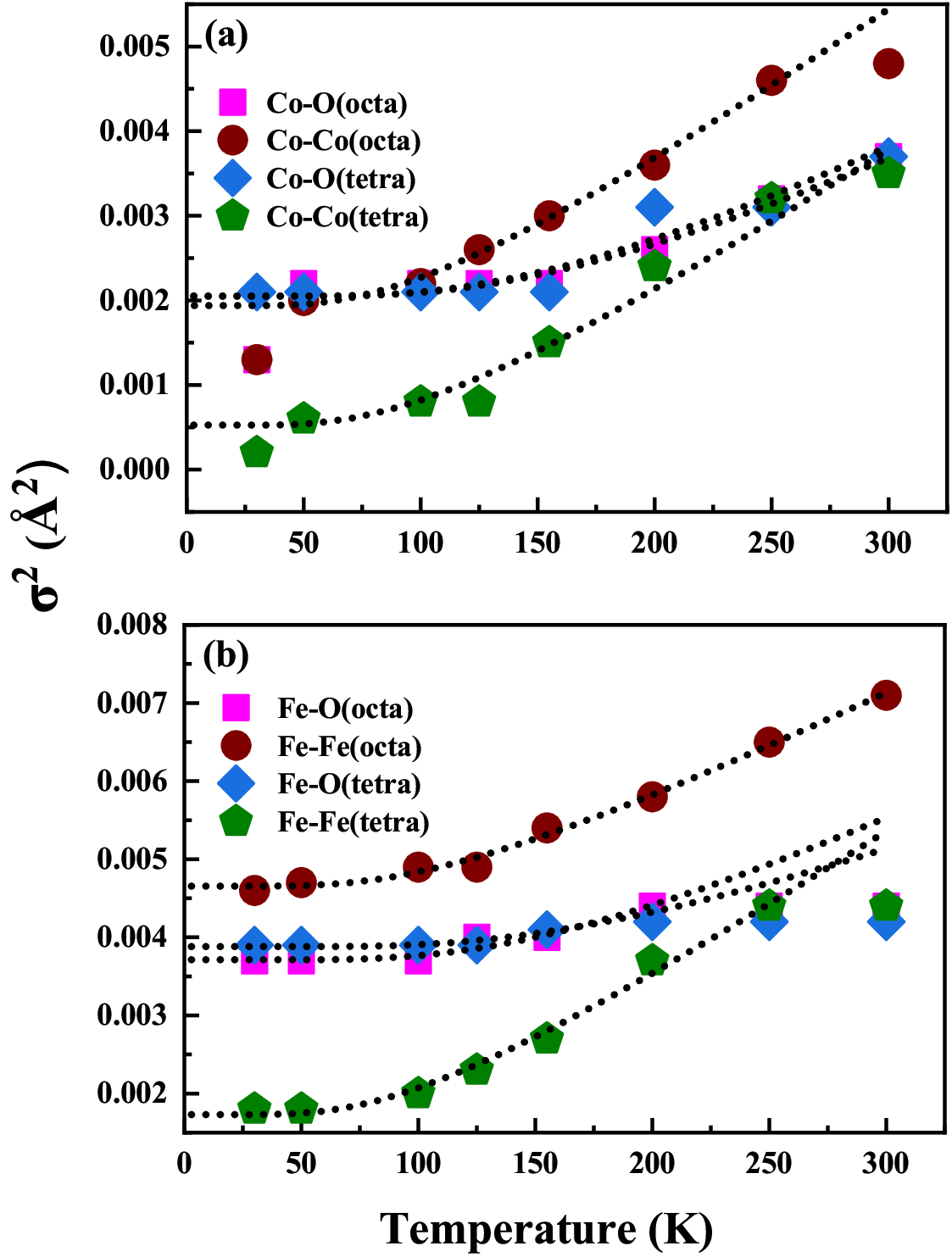}
\caption{\label{fig:wide} \texttt{}Debye-Waller factor from EXAFS analysis for CFO samples. The dashed lines are the best-fit data obtained using a correlated Einstein model (a) Co-K edge, (b) Fe-K edge.}
\end{figure}

Again, to evaluate the oxidation states of the Co and Fe cations in CoFe$_2$O$_4$, bond valence sum (BVS) calculations were performed using the inbuilt module in the FullProf Suite~\cite{rodriguez1993recent}. The BVS method estimates the valence state \( V \) of a cation based on its local bonding environment using the expression \( V = \sum_i \exp\left( \frac{d_0 - d_i}{0.37} \right) \), where \( d_i \) represents the experimentally determined bond lengths to surrounding anions, and \( d_0 \) is a tabulated empirical parameter characteristic of the specific cation-anion pair~\cite{brese1991bond,brown2016chemical}. Therefore, the local distortion from regular tetrahedral and octahedral (O$_h$) environment have been investigated by calculating the bond-valance sum (BVS) of Co/Fe-O in Co/FeO$_4$ tetrahedral and Co/FeO$_6$ octahedra environment for different temperatures across the structural (across the super-lattice peak) phase transition. We have not observed much change above the super-structure phase transition i.e. 200 K. However, below 200 K, we have found pronounced changes in Co/Fe-O bond distance indicating main role of lattice distortion due to phase-transitions, as is clearly reflected in BVS as a function of temperature, (shown in Fig. 8 (a, b)). These results clearly suggest that distortion led bond-length variation as a function of temperature across the super-structure phase-transition shows strong lattice-coupling, which leads to symmetry-breaking in CoFe$_2$O$_4$. Thus, resulting BVS values exhibit clear temperature-dependent variations in the effective charge states of both Co and Fe. These shifts become more pronounced in the intermediate temperature range (\( T_o \approx 200\text{--}100\,\text{K} \)), suggesting dynamic charge redistribution between Co$^{2+}$/Co$^{3+}$ and Fe$^{2+}$/Fe$^{3+}$. Such fluctuations may reflect short-range charge ordering or charge-orbital correlations \cite{yogi2022coexisting} shown in Fig. 8, which are consistent with the structural distortions, phonon anomalies, and hysterqesis in the temperature-dependent magnetic behavior observed in the same temperature regime.

\subsection{X-ray absorption fine structure spectroscopy }
\subsubsection{X-ray absorption near edge spectroscopy (XANES)}

The XAFS signal provides information about interatomic distances, atomic structure, and the electrical nature of the absorber. It is typically divided into the EXAFS (extended) and XANES (near-edge) regions. The normalized Co and Fe K-edge are presented in Fig.9  to probe the electronic structure around both Fe and Co atoms. The XANES region provides rich structural and electronic information and is generally weakly temperature-dependent due to the Debye-Waller factor ($e^{-2k^2\sigma^2}$) \cite{bera2024local,benfatto2014close}. The edge position reflects the valence state of the photoabsorber. With the extended mean free path of low-energy photoelectrons, multiple scattering effects significantly influence the XANES features, making them highly sensitive to the coordination symmetry around the photoabsorber. Additionally, pre-edge analysis offers valuable insights, particularly for the K-edges of 3$d$ metals.The Co and Fe K-edge XANES spectra were processed using standard methods, including pre-edge removal, post-edge atomic background subtraction ($\alpha_0$), and edge jump normalization \cite{meneghini2012estra,koningsberger1987x}. The edge energy ($E_0$), defining the origin of the photoelectron wave vector $k = \hbar^{-1} \sqrt{2m_e(E - E_0)}$ ($m_e$ being the electron mass), was determined at the midpoint of the edge rise and refined during analysis.

The Co and Fe K-edge XANES spectra show similar behavior with respect to temperature, with the standard deviation between the normalized spectra being less than \( 10^{-3} \). The pre-edge feature is due to electronic transitions from the 1s core level of the absorbing atom (e.g., Co or Fe) to unoccupied d states. The intensity of this pre-edge feature increases with the degree of p-d hybridization \cite{yamamoto2008assignment, wu2004quadrupolar, probingthe-electronic-SrLaCoNbO6}. Here we note that the low-intensity pre-edge features in the XANES spectra of transition metal oxides arise from 1s electron transitions to unoccupied d states (quadrupole transitions, $\Delta l = 2$) or to d states hybridized with p states (metal or oxygen 2p) \cite{yamamoto2008assignment,probingthe-electronic-SrLaCoNbO6,glatzel2009resonant}.According to group theory, ideal octahedral symmetry (\(O_h\)) prevents p-d mixing, leading to the absence of pre-edge features. However, structural distortions break this symmetry, allowing p-d hybridization and enhancing the pre-edge intensity, which becomes more pronounced with increasing off-center displacement of the absorbing atoms.\cite{yamamoto2008assignment,probingthe-electronic-SrLaCoNbO6}.For the Co K-edge XANES spectra of CoFe$_2$O$_4$, no significant pre-edge intensity is observed near 7708.40 eV, whereas for Fe, a pre-edge feature is clearly distinct around 7111 eV. This difference indicates a more pronounced pre-edge feature for Fe in the FeO$_6$ octahedra and the weak pre-edge feature for Co may also result from thermally induced dynamic distortions in the CoO$_6$ octahedra, observable at both room and low temperatures. The pre-edge feature in both edges is denoted as 'a' in Fig. 9(a, b) and marked with (*) in Fig. 9(c, d). The rising edge (approximately 7717 eV for Co and 7121 eV for Fe) and other distinct features (labeled  b, c, d, e, f) in both edge spectra [see Fig. 9 (a,b)] have been analyzed. A shift of approximately 1~eV in the rising edge of the XANES spectra is observed upon cooling, while the white line remains nearly unchanged, indicating no significant change in the oxidation state of Co and Fe. This shift is attributed to Jahn-Teller distortion at low temperature, which leads to elongation of Co/Fe-O bonds. As previously reported,\cite{natoli1984distance,ramos701759stability} such bond-length variations can induce shift in the edge positions, often exceeding those caused by initial-state effects due to electronic configuration changes, which are typically less than 1~eV. This behaviour is consistent with ligand field theory, wherein back-donation from ligands partially compensates for charge transfer, resulting in minimal changes to the effective charge on the metal ions\cite{natoli1984distance,ramos701759stability}. Comparisons with reference samples, such as Co$_3$O$_4$ and CoO for Co, and FeO and Fe$_2$O$_3$ for Fe, confirm the mixed valence state of Co and Fe across all temperature ranges.

The minor feature labeled 'b', observed around 7717 eV and 7121 eV, respectively (shown in fig-9 (a,b)) for Co and Fe-edge, respectively, is attributed to charge transfer from oxygen to Co or Fe ions in the final state, which results from the strong hybridization between Co or Fe 3d and O 2p orbitals. This suggests a small charge transfer energy in these compounds. The white line for both Co and Fe K-edges is observed around 7725 eV and 7126 eV, respectively, and is primarily attributed to the \(1s \rightarrow 4p\) dipole transition. The position of the white line is influenced by both the valence state and the local coordination of the absorbing atom \cite{ishimatsu2015differences}. Octahedral site disorder affects the main-edge feature similarly for both Co and Fe K-edge spectra, indicating that the white line intensity is mainly controlled by the level of octahedral distortion. In the temperature range associated with orbital ordering, no noticeable change was observed in the white line intensity. The post-edge features (E, F) observed in the Co and Fe K-edge XANES spectra of CoFe$_2$O$_4$ correspond to electronic transitions in which 1s electrons from Co or Fe are excited into Co 4p or Fe 4p states. These 4p states are strongly hybridized with O 2p orbitals, reflecting the interaction between cobalt, iron, and oxygen in the lattice.

The linear combination fit (LCF) analysis of the XANES of the reference file yielded consistent results with the HXPES data. For cobalt, the analysis indicated that the sample contains approximately 65\% Co\textsuperscript{2+} and 35\% Co\textsuperscript{3+}, while for iron, it revealed a distribution of around 45\% Fe\textsuperscript{2+} and 55\% Fe\textsuperscript{3+} at room temperature. This analysis suggests that a fraction of the \(\text{Co}^{2+}\) ions may have migrated to the A site and oxidized to \(\text{Co}^{3+}\), while a portion of the \(\text{Fe}^{3+}\) ions are originally at the A site shifted to the B site and reduced to \(\text{Fe}^{2+}\), preserving charge neutrality. This implies the formation of antisite defects, where Co ions move from the B site to the A site, while Fe ions at the A site relocate to the B site \cite{kobayashi2022intervalence}.

\subsubsection{Extended X-ray absorption fine structure (EXAFS)}

Before discussing the results of EXAFS, it is important to highlight the complementary nature of XAFS and XRD techniques. XRD captures structural features with periodic repetition in unit cells, while non-periodic features contribute only to a diffuse background. In contrast, XAFS focuses on the local structure around the absorber, regardless of short- or long-range order. This distinction highlights the unique and complementary insights provided by the two methods \cite{bera2024local}.Generally, nonperiodic features in the atomic distribution can cause the interatomic distances obtained from EXAFS analysis to differ significantly from those between atomic sites in the periodic lattice.

 In the EXAFS region, the signal is dominated by single photoelectron scattering (SS) and selected multiple scattering (MS) processes from nearly collinear atomic configurations, with enhancements from forward scattering effects \cite{filipponi1995x,bunker2010introduction}. The EXAFS signal is typically modeled using the standard EXAFS formula, which describes the average local structure around the absorber as a sum of Gaussian-shaped interatomic shells. Each shell is characterized by its multiplicity (N), average distance (R), and mean square relative displacement (MSRD, $\sigma^2$) \cite{bunker2010introduction}. The $\sigma^2$ consists of a temperature-dependent component, which accounts for thermally activated lattice vibrations, and a temperature-independent static component:

\begin{equation}
\sigma^2 = \sigma^2_{\text{E}}(T) + \sigma^2_{\text{static}}.
\end{equation}
The temperature-dependent term is accurately described by a correlated Einstein model \cite{vaccari2006einstein}.
\begin{equation}
\sigma_E^2(T) = \frac{\hbar}{\mu \omega_E} \coth\left(\frac{\hbar \omega_E}{2 k_B T}\right).
\end{equation}

Note that at low temperatures (T less than approximately 60–70 K),$\sigma^2_{\text{E}}(T)$ typically remains nearly constant\cite{bunker2010introduction,filipponi1995x}.By ensuring the stability and reproducibility of the experimental setup and minimizing correlations between fitting parameters (e.g., by fixing $\Delta E$ and coordination numbers), the accuracy in determining relative variations in structural parameters as a function of external factors like temperature can be improved significantly, as demonstrated in studies of thermal expansion in solids \cite{vaccari2006einstein}. While analyzing next-neighbor shells, accuracy may decrease due to signal overlap, parameter correlations, and the need to merge or neglect certain contributions. However, the relative variations in parameters can reveal even subtle structural effects.

Fig.~10 shows the temperature-dependent EXAFS data at the (a) Co K-edge and (b) Fe K-edge. The $y$-axis represents the magnitude of the Fourier-transformed EXAFS function, $|\chi(R)|$ (in \AA$^{-3}$), while the $x$-axis shows the radial distance $R$ (in \AA) from the absorbing atom, corresponding to the real-space positions of neighboring atoms, revealing an increase in local structural disorder with rising temperature. At both edges, the amplitudes of the \( \chi(R) \) peaks decrease with temperature in both coordination shells, indicating thermal vibrations.

\begin{table}[t]
\centering
\begin{tabular}{ccccc}
\hline
\textbf{T (K)} & \textbf{Path} & \textbf{N} & \textbf{R (\AA)} & \textbf{$\sigma^2$ (\AA$^2$)} \\
\hline
\textbf{\multirow{4}{*}{300K}} & Co-O (Octa) & 6 & 2.103 ± 0.002 & 0.0037 ± 0.0002 \\
 & Co-Co (Octa) & 6 & 3.008 ± 0.003 & 0.0048 ± 0.0001 \\
 & Co-O (Tetra) & 4 & 1.916 ± 0.002 & 0.0037 ± 0.0003 \\
 & Co-Co (Tetra) & 4 & 3.491 ± 0.003 & 0.0035 ± 0.0003 \\
\hline
\textbf{\multirow{4}{*}{250K}} & Co-O (Octa) & 6 & 2.103 ± 0.002 & 0.0032 ± 0.0002 \\
 & Co-Co (Octa) & 6 & 3.008 ± 0.003 & 0.0046 ± 0.0003 \\
 & Co-O (Tetra) & 4 & 1.919 ± 0.002 & 0.0031 ± 0.0003 \\
 & Co-Co (Tetra) & 4 & 3.491 ± 0.003 & 0.0032 ± 0.0003 \\
\hline
\textbf{\multirow{4}{*}{200K}} & Co-O (Octa) & 6 & 2.103 ± 0.002 & 0.0026 ± 0.0002 \\
 & Co-Co (Octa) & 6 & 3.008 ± 0.003 & 0.0036 ± 0.0003 \\
 & Co-O (Tetra) & 4 & 1.919 ± 0.002 & 0.0031 ± 0.0003 \\
 & Co-Co (Tetra) & 4 & 3.491 ± 0.003 & 0.0024 ± 0.0003 \\
\hline
\textbf{\multirow{4}{*}{155K}} & Co-O1 (Octa) & 4 & 2.078 ± 0.002 & 0.0022 ± 0.0002 \\
 & Co-O2 (Octa) & 2 & 2.41 ± 0.002 & 0.0022 ± 0.0002 \\
 & Co-Co (Octa) & 6 & 3.008 ± 0.003 & 0.0030 ± 0.0003 \\
 & Co-O (Tetra) & 4 & 1.919 ± 0.002 & 0.0021 ± 0.0003 \\
 & Co-Co (Tetra) & 4 & 3.491 ± 0.003 & 0.0015 ± 0.0003 \\
\hline
\textbf{\multirow{4}{*}{125K}} & Co-O1 (Octa) & 4 & 2.084 ± 0.002 & 0.0022 ± 0.0002 \\
 & Co-O2 (Octa) & 2 & 2.126 ± 0.002 & 0.0022 ± 0.0002 \\
 & Co-Co (Octa) & 6 & 3.008 ± 0.003 & 0.0026 ± 0.0003 \\
 & Co-O (Tetra) & 4 & 1.919 ± 0.002 & 0.0021 ± 0.0003 \\
 & Co-Co (Tetra) & 4 & 3.491 ± 0.003 & 0.0008 ± 0.0003 \\
\hline
\textbf{\multirow{4}{*}{100K}} & Co-O1 (Octa) & 4 & 2.092 ± 0.002 & 0.0022 ± 0.0002 \\
 & Co-O2 (Octa) & 2 & 2.132 ± 0.002 & 0.0022 ± 0.0002 \\
 & Co-Co (Octa) & 6 & 3.008 ± 0.003 & 0.0022 ± 0.0003 \\
 & Co-O (Tetra) & 4 & 1.919 ± 0.002 & 0.0021 ± 0.0003 \\
 & Co-Co (Tetra) & 4 & 3.491 ± 0.003 & 0.0008 ± 0.0003 \\
\hline
\textbf{\multirow{4}{*}{50K}} & Co-O (Octa) & 6 & 2.103 ± 0.002 & 0.0022 ± 0.0002 \\
 & Co-Co (Octa) & 6 & 3.008 ± 0.003 & 0.0020 ± 0.0003 \\
 & Co-O (Tetra) & 4 & 1.919 ± 0.002 & 0.0021 ± 0.0003 \\
 & Co-Co (Tetra) & 4 & 3.491 ± 0.003 & 0.0006 ± 0.0003 \\
\hline
\textbf{\multirow{4}{*}{30K}} & Co-O (Octa) & 6 & 2.103 ± 0.002 & 0.0013 ± 0.0002 \\
 & Co-Co (Octa) & 6 & 3.008 ± 0.003 & 0.0013 ± 0.0003 \\
 & Co-O (Tetra) & 4 & 1.919 ± 0.002 & 0.0021 ± 0.0003 \\
 & Co-Co (Tetra) & 4 & 3.491 ± 0.003 & 0.0002 ± 0.0003 \\
\hline
\end{tabular}

\caption{Co K-edge EXAFS data fitting, where CN (coordination number) was kept as a fixed parameter, and R (bond distance) and $\sigma^2$ (Debye-Waller factor) were variable parameters. The numbers in parentheses indicate the uncertainty in the last digit.}

\label{table:CoEdge}
\end{table}

\begin{table}[h!]
\centering
\begin{tabular}{ccccc}
\hline
\textbf{T (K)} & \textbf{Path} & \textbf{N} & \textbf{R (\AA)} & \textbf{$\sigma^2$ (\AA$^2$)} \\
\hline
\textbf{\multirow{4}{*}{300K}} & Fe-O (Octa) & 6 & 1.998 ± 0.003 & 0.0044 ± 0.0001 \\
 & Fe-Fe (Octa) & 6 & 3.038 ± 0.003 & 0.0071 ± 0.0001 \\
 & Fe-O (Tetra) & 4 & 1.839 ± 0.003 & 0.0042 ± 0.0001 \\
 & Fe-Fe (Tetra) & 4 & 3.523 ± 0.003 & 0.0044 ± 0.0001 \\
\hline
\textbf{\multirow{4}{*}{250K}} & Fe-O (Octa) & 6 & 1.998 ± 0.003 & 0.0044 ± 0.0001 \\
 & Fe-Fe (Octa) & 6 & 3.038 ± 0.003 & 0.0065 ± 0.0001 \\
 & Fe-O (Tetra) & 4 & 1.839 ± 0.003 & 0.0042 ± 0.0001 \\
 & Fe-Fe (Tetra) & 4 & 3.523 ± 0.003 & 0.0044 ± 0.0001 \\
\hline
\textbf{\multirow{4}{*}{200K}} & Fe-O (Octa) & 6 & 1.998 ± 0.003 & 0.0044 ± 0.0001 \\
 & Fe-Fe (Octa) & 6 & 3.038 ± 0.003 & 0.0058 ± 0.0001 \\
 & Fe-O (Tetra) & 4 & 1.839 ± 0.003 & 0.0042 ± 0.0001 \\
 & Fe-Fe (Tetra) & 4 & 3.523 ± 0.003 & 0.0037 ± 0.0001 \\
\hline
\textbf{\multirow{4}{*}{155K}} & Fe-O (Octa) & 4 & 1.998 ± 0.003 & 0.0040 ± 0.0001 \\
 & Fe-O (Octa) & 2 & 2.005 ± 0.003 & 0.0040 ± 0.0001 \\
 & Fe-Fe (Octa) & 6 & 3.038 ± 0.003 & 0.0054 ± 0.0001 \\
 & Fe-O (Tetra) & 4 & 1.839 ± 0.003 & 0.0041 ± 0.0001 \\
 & Fe-Fe (Tetra) & 4 & 3.523 ± 0.003 & 0.0027 ± 0.0001 \\
\hline
\textbf{\multirow{4}{*}{125K}} & Fe-O1 (Octa) & 4 & 1.998 ± 0.003 & 0.0040 ± 0.0001 \\
 & Fe-O2 (Octa) & 2 & 2.005 ± 0.003 & 0.0040 ± 0.0001 \\
 & Fe-Fe (Octa) & 6 & 3.038 ± 0.003 & 0.0049 ± 0.0001 \\
 & Fe-O (Tetra) & 4 & 1.839 ± 0.003 & 0.0039 ± 0.0001 \\
 & Fe-Fe (Tetra) & 4 & 3.523 ± 0.003 & 0.0023 ± 0.0001 \\
\hline
\textbf{\multirow{4}{*}{100K}} & Fe-O1 (Octa) & 4 & 1.999 ± 0.003 & 0.0039 ± 0.0001 \\
 & Fe-O2 (Octa) & 2 & 2.004 ± 0.003 & 0.0038 ± 0.0001 \\
 & Fe-Fe (Octa) & 6 & 3.038 ± 0.003 & 0.0049 ± 0.0001 \\
 & Fe-O (Tetra) & 4 & 1.839 ± 0.003 & 0.0039 ± 0.0001 \\
 & Fe-Fe (Tetra) & 4 & 3.523 ± 0.003 & 0.0020 ± 0.0001 \\
\hline
\textbf{\multirow{4}{*}{50K}} & Fe-O (Octa) & 6 & 1.998 ± 0.003 & 0.0037 ± 0.0001 \\
 & Fe-Fe (Octa) & 6 & 3.038 ± 0.003 & 0.0047 ± 0.0001 \\
 & Fe-O (Tetra) & 4 & 1.839 ± 0.003 & 0.0039 ± 0.0001 \\
 & Fe-Fe (Tetra) & 4 & 3.523 ± 0.003 & 0.0018 ± 0.0001 \\
\hline
\textbf{\multirow{4}{*}{30K}} & Fe-O (Octa) & 6 & 1.998 ± 0.003 & 0.0037 ± 0.0001 \\
 & Fe-Fe (Octa) & 6 & 3.038 ± 0.003 & 0.0046 ± 0.0001 \\
 & Fe-O (Tetra) & 4 & 1.839 ± 0.003 & 0.0039 ± 0.0001 \\
 & Fe-Fe (Tetra) & 4 & 3.523 ± 0.003 & 0.0018 ± 0.0001 \\
\hline
\end{tabular}

\caption{Fe K-edge EXAFS data fitting, where CN (coordination number) was kept as a fixed parameter, and R (bond distance) and $\sigma^2$ (Debye-Waller factor) were variable parameters. The numbers in parentheses indicate the uncertainty in the last digit.}
\label{table:FeEdge}
\end{table}

Figures 11 and 12 show the fitting of temperature-dependent EXAFS data for both Co and Fe edges across all temperatures. The corresponding fitting parameters are provided in Tables 1 and 2. The findings from the fitting analysis are summarised below. The results highlight key temperature-dependent trends in local structural parameters, particularly in the Co–O and Fe–O bond lengths. At higher temperatures (above 155\,K), the Co–O and Fe–O bond lengths in octahedral coordination remain nearly uniform, consistent with an undistorted local environment. However, a clear bond length splitting is observed at or below 155\,K, 125\,K, and 100\,K for both Co–O and Fe–O octahedral paths. This behaviour is a hallmark of Jahn–Teller distortion (JTD), \cite{koborinai2016orbital} in octahedrally coordinated Co$^{2+}$ and Fe$^{3+}$ ions [Fig. 1] and leads to anisotropic elongation of the surrounding oxygen bonds which modify the spin ground state of Co$^{2+}$ and Fe$^{3+}$ ions due to crystal electric field (CEF) [Fig. 1]. This separation vanishes again at lower temperatures (50\,K and 30\,K), where the bond lengths return to near uniform values, suggesting a suppression or freezing of dynamic distortions.  In addition, from the temperature dependent SR-XRD along with EXAFS as a local probe measurements, we have observed evidence of super-structure, which occurs due to significant local distortion of CoO4/FeO4 and CoO6/FeO6 at structural transition ($T_o$ = 200 K), strongly influencing the spins associated with Co$^{2+}$ and Fe$^{3+}$.The consequence of this structural transition is orbital corelation transition in CFO. Importantly, significant change in the bond length is not observed in the tetrahedral coordination shells, indicating that the Jahn–Teller distortion is confined to the octahedral sublattice [please see, Fig. 1]. The emergence of JTD in the intermediate temperature range points to a local structural instability, possibly induced by short-range orbital \cite{tokura2000orbital,koborinai2016orbital} ordering or spin–orbital coupling.

The temperature dependence of the Debye-Waller factor (\(\sigma^2\)) provides critical insights into the local structural dynamics and the extent of disorder within the studied sample. In the present work, the \(\sigma^2\) values for Fe-O and Fe-Fe bonds in octahedral coordination exhibit a notable increase with temperature, highlighting the contributions of both thermal vibrations and static disorder. The presence of oxygen vacancies contributes to enhanced lattice instability, as reflected in the progressive rise of \(\sigma^2\). For the tetrahedral coordination sites, the \(\sigma^2\) values of Fe-O and Fe-Fe bonds remain relatively constant and lower compared to the octahedral coordination sites. This indicates that the tetrahedral framework is more resistant to disorder and less impacted by the redistribution of cations or the introduction of oxygen vacancies. Such structural robustness suggests a localized decoupling between the dynamic behavior of tetrahedral and octahedral sites, with the latter being more susceptible to distortions. The octahedral \(\sigma^2\) values, particularly for Co-O and Co-Co bonds, exhibit a pronounced temperature-dependent rise, underscoring the pivotal role of cation redistribution and oxygen deficiency in generating disorder. This distinction between octahedral and tetrahedral behavior reflects the asymmetric distribution of strain and local lattice distortions in the sample, primarily concentrated in the octahedral sublattice. The rapid increase in \(\sigma^2\) for Fe-Fe and Co-Co octahedral bonds highlights the impact of these defects on the structural integrity of the material. 

From a broader perspective, the differences in the behavior of \(\sigma^2\) between tetrahedral and octahedral sites have significant implications for the material’s electronic and magnetic properties. The increased disorder in octahedral sites may lead to localized changes in the electronic structure, such as alterations in the crystal field splitting energy, which in turn can influence the orbital occupancy and electronic transitions of Fe and Co ions. These distortions can also affect the magnetic exchange \cite{kugel1973crystal} interactions, potentially introducing frustration or disorder into the magnetic ground state.

\begin{table}[h!]
\centering
\caption{Fitted Parameters: Einstein Temperature (\(\theta_E\)) and Static Disorder (\(\sigma_S^2\))}
\begin{tabular}{lcc}
\hline
\textbf{Bond Type} & \(\theta_E\) (K) & \(\sigma_S^2\) (\(\mathrm{\AA^2}\)) \\
\hline
Fe-Fe (octa)       & $337.16 \pm 4.07$     & $2.08 \times 10^{-4} \pm 6.64 \times 10^{-5}$ \\
Fe-Fe (tetra)      & $291.21 \pm 4.67$     & $1.24 \times 10^{-4} \pm 1.01 \times 10^{-5}$ \\
Fe-O (octa)        & $650.91 \pm 49.68$    & $8.34 \times 10^{-4} \pm 2.96 \times 10^{-5}$ \\
Fe-O (tetra)       & $605.63 \pm 23.08$    & $6.71 \times 10^{-4} \pm 1.39 \times 10^{-5}$ \\
\hline
Co-O (octa)        & $496.17 \pm 27.24$    & $1.98 \times 10^{-4} \pm 3.31 \times 10^{-5}$ \\
Co-O (tetra)       & $510.45 \pm 20.40$    & $1.72 \times 10^{-4} \pm 2.29 \times 10^{-5}$ \\
Co-Co (octa)       & $289.64 \pm 3.64$     & $9.12 \times 10^{-4} \pm 8.06 \times 10^{-5}$ \\
Co-Co (tetra)      & $296.30 \pm 13.08$    & $2.29 \times 10^{-4} \pm 3.34 \times 10^{-5}$ \\
\hline
\end{tabular}
\end{table}

Using the Einstein model from Eq.~(2), we fitted the Debye-Waller factor, and the results are presented in Fig.~13 for each coordination environment. We observe Fe-Fe octa site shown in Fig.~13(b), the highest disorder. The primary cause of disorder at the Fe sites may be attributed to crystal field splitting ($\Delta_{\mathrm{CFS}}$), which significantly influences Fe–Fe interactions, particularly in octahedral environments. In octahedral coordination, Fe$^{3+}$ ($d^5$) and Fe$^{2+}$ ($d^6$) ions experience stronger crystal field splitting ($\Delta_{\mathrm{oct}}$), which weakens direct Fe–Fe interactions. This reduced interaction strength results in increased flexibility of the Fe sites, making them more susceptible to local structural disorder. In contrast, Fe–Fe interactions within tetrahedral coordination environments experience a much lower crystal field splitting ($\Delta_{\mathrm{tet}} < \Delta_{\mathrm{oct}}$), leading to comparatively greater structural stability. A similar trend is observed for Co sites; however, due to the weaker Co–O covalency relative to Fe–O, the octahedral Co–Co disorder is less pronounced than that observed in the Fe–Fe (octahedral) network.

\section{Conclusion}

The structural and electronic disorder in CoFe$_2$O$_4$ (CFO), revealed through synchrotron XRD, EXAFS, and XANES, has a pronounced influence on the spectral characteristics observed in HAXPES. Cation redistribution and oxygen vacancies contribute to lattice disorder and local strain within the coordination environment. These effects manifest in HAXPES as broadened and asymmetric Co and Fe core-level spectra, indicative of mixed valence states. The presence of strong shake-up satellite features reflects enhanced charge-transfer excitations and electronic correlations arising from the disordered local structure.

Temperature-dependent EXAFS analysis, through the Debye--Waller factor, confirms an increase in thermal and static disorder, which further impacts the spectral line shapes and multiplet splitting observed in HAXPES. The disruption of local crystal field environments by oxygen vacancies modifies screening conditions and charge-transfer states, highlighting the close coupling between structural distortions and electronic behavior. These findings are reinforced by Raman spectroscopy, which reveals significant temperature-dependent shifts and broadening of A$_{1g}$, E$_g$, and T$_{2g}$ modes. The activation of additional Raman modes at lower temperatures, along with phonon anomalies, points to magnetoelastic coupling and dynamic lattice effects.

Magnetic measurements exhibit a clear thermal hysteresis between field-cooled cooling (FCC) and warming (FCW) curves, suggesting competition between tetrahedral and octahedral magnetic sublattices. As temperature decreases, antiferromagnetic and ferromagnetic interactions become increasingly competitive, modifying the local magnetic environment. These changes are reflected in the Raman spectral response, underscoring the strong interplay between lattice, spin, and electronic degrees of freedom.

In conclusion, the detailed temperature dependent structural investigation using powder synchrotron x-ray diffraction and extracted refinement parameters such as unit-cell parameters and bond-lengths as a function of temperature suggests presence of strong distortion, further confirmed from local probe EXAFS. Accordingly, strong lattice-coupling gradually develops orbital correlations in CoFe$_2$O$_4$. These correlations become significant below 200 K, where an emergence of super-lattice peak could remarkably induce symmetry-breaking. This is because of local octahedral and tetrahedral environment distortion, most likely responsible for the origin of orbital order formation in CFO. Around the orbital transition temperature ($T_o \approx 200$--$100$~K), we observe clear signatures of orbital ordering associated with cooperative Jahn--Teller distortions, spin reorientation or evolving exchange interactions, and charge redistribution, as supported by bond valence sum (BVS) analysis. Below $T_o$, all associated structural, magnetic, and vibrational anomalies vanish, indicating the onset of a frozen, disordered orbital configuration consistent with an orbital glass state. These findings offer new insights into the frustrated coupling of orbital, charge, spin, and lattice sectors in spinel ferrites.



\begin{acknowledgments}

\section{Acknowledgments}
The authors express their sincere gratitude to M.P. Saravanan and the cryogenics team for their assistance in supplying cryogens. We acknowledge the University Grants Commission (UGC), Government of India, for the financial support through a fellowship grant. The authors also thank Pragati Sharma for performing the Raman measurements.Special thanks are due to Priyanshi Tiwari, Rajeev Joshi, Pramod R. Nadig, and Ananya Sahoo or their insightful discussions and valuable suggestions.

\end{acknowledgments}

\appendix

\nocite{*}
\bibliography{CFO}
\bibliographystyle{unsrt}

\end{document}